%% file: main.tex
\documentclass[apj,twocolumn]{openjournal}
\usepackage{amsmath,amstext} 
\usepackage[T1]{fontenc}
\usepackage{apjfonts} 
\usepackage{longtable}
\usepackage{comment}

\begin{document}
\include{macros}
\title[RUBIES Ionization Parameter]{RUBIES: The Evolution of the Ionization Parameter from 0 < z < 9\vspace{-15mm}}
\include{authors}

\begin{abstract}
% Here is what we want
    The dimensionless ionization parameter, $U\equiv q/c$, where $q$ is the ratio of the local ionizing photon flux to the local hydrogen density, is a key metric to parameterize nebular conditions. 
% Here is the problem/setting up the tension
    Prior to JWST, the rest-frame optical emission lines and their ratios which trace the ionization parameter (e.g., O32$\equiv$\oiii/\oii) were inaccessible at high redshifts.
% Here is the novelty
    Here we quantify, for the first time, the evolution of the ionization parameter in galaxies across the last 13 billion years of cosmic time by comparing JWST/NIRSpec PRISM and G395M spectroscopy of 434 galaxies at $3\lesssim z\lesssim9$ from the RUBIES survey with $z<3$ samples from SDSS, LEGA-C, and KBSS. 
% Here is what we did
    We leverage a large suite of Cloudy photoionization models to infer the ionization parameter from \oiii\ and \oii. 
% Here is what we found
% New Result
    We find that U increases with redshift and specific star formation rate (sSFR), and decreases with stellar mass. 
% Comparison to old results
    Crucially, and in contrast to previous linear best-fit calibrations, our inference results in a systematic uncertainty in log U of $\sim0.3$ dex at zero measurement uncertainty due to the wide range of photoionization models that predict the same O32 ratio without informative priors. 
% Comparison to theory
    We compare to \sphinx\ and LUMEN simulations and find that the simulated galaxies exhibit higher O32 ratios at fixed redshift and stellar mass compared to RUBIES observations.
% New application of results
    Finally, we combine the predictive power of observed and physical quantities with multivariate relations to estimate U from redshift, stellar mass, and sSFR for use in cases where O32 is not available from spectroscopy. 
% Here is why the key result is important
    We find that the ionization parameter increases at fixed stellar mass and sSFR by a factor of $\sim4$ from $z=2$ to $z=6$, demonstrating that the redshift evolution encapsulates physics beyond that traced by stellar mass and sSFR alone. Finally, we show that a toy model with the first order assumption that H II region volume is proportional to galaxy volume can explain the excess redshift dependence of the ionization parameter as being consistent with observed evolution in galaxy sizes. 

\end{abstract}

\section{Introduction}\label{sec:introduction}
Spectroscopy of strong rest-frame UV/optical emission lines can be used to determine a wide range of physical properties of the observed galaxy, from determining sources of ionizing photons \citep[e.g., ``BPT''-style diagrams][]{Baldwin1981,Veilleux1987,Kewley2006,Kewley2019b,Trump2013,Trump2015}, to deriving the conditions of the interstellar medium (ISM) such as gas-phase metallicities, temperatures, and densities \citep[e.g.,][]{Pagel1979,Zaritsky1994,Wilson1994,Kewley2002,Kewley2008,Tremonti2004,Erb2006,Mannucci2010,Maiolino2019,Sanders2016,Sanders2020,Steidel2014,Steidel2016,Strom2017,Strom2018,Shapley2015,Shapley2019,Shapley2025,Topping2020a,Topping2020b,Cullen2021,Runco2021}, to estimating star formation rates \citep[e.g.,][]{Kennicutt1998a,Kennicutt1998b,Kennicutt2012,Kewley2004} and dust attenuation \citep[e.g.,][]{Calzetti1994,Calzetti2000,Salim2020}. 

The ionization parameter is a key metric to quantify nebular conditions, though it is rarely the direct focus of high-redshift galaxy studies due to the intrinsic degeneracies of the conversion from observed emission line ratios (e.g., O32) to the ionization parameter \citep[e.g.,][and see Section \ref{sec:discussion:logU_methodology} for discussion]{Sanders2023}. The ionization parameter is defined as
\begin{align}\label{eq:ionization_parameter}
    q &\equiv \frac{\Phi_{H^0}}{n_H}
\end{align}
where $\Phi_{H^0}$ is the local ionizing photon flux incident on the face of the cloud and $n_H$ is the local hydrogen density. This ionization parameter has units of cm s$^{-1}$, and can be interpreted to first order as the speed at which the ionization front expands into the surrounding neutral medium \citep[e.g.,][]{Kewley2019b}. Throughout this work, we refer to the dimensionless ionization parameter, $U\equiv q/c$. 

To constrain the ionization parameter through observations, studies often use ratios of strong optical emission lines which trace different ``ionization zones'': regions of a nebula where ions of similar ionization energies are present \citep[e.g.,][]{Strom2017,Strom2018,Kewley2002,Kewley2019b,Berg2021,Papovich2022}. The four-zone model \citep{Berg2021} defines the zones by the ionization energies needed to produce N$^+$ (``low'': 14.53-29.60 eV), S$^{2+}$ (``intermediate'': 23.33-34.79 eV), O$^{2+}$ (``high'': 35.12-54.94 eV), and He$^{2+}$ (``very high'': >54.42 eV). One such line ratio is O32\footnote{Some other works may define the O32 line ratio as $\mathrm{O32} \equiv \frac{\oiii~\lambda\lambda4960,5008}{\oii~\lambda\lambda3727,3730}$, though each convention can be rescaled to match the other through the fixed $\oiii~\lambda5008/\oiii\lambda4960 = 2.985$ \citep[e.g.,][]{Storey2000}}, where
\begin{align}\label{eq:O32}
    \mathrm{O32} &\equiv \frac{\oiii~\lambda5008}{\oii~\lambda\lambda3727,3730}
\end{align}
The \oiii\ emission lines arising from doubly-ionized oxygen (\Otwoplus) probe ionization energies 35.1-54.9 eV, and \oii\ emission lines from singly-ionized oxygen (\Oplus) require 13.6-35.1 eV.

The O32 ratio represents an ``intermediate'' ionization parameter as a ratio between a ``high'' and ``low'' ionization species \citep[e.g.,][]{Kewley2002,Sanders2016,Berg2021}, which is the energy regime traced by the most common abundance, density, and ionizing source diagnostics. A primary advantage of O32 is the insensitivity to relative elemental abundances compared to other similar-ionization strong line ratios accessible in the rest-frame optical, such as $\oiii~\lambda5008$/\hb\ and $\neiii~\lambda3870$/\oii\ \citep[e.g.,][]{Berg2021}. O32 is observable in JWST/NIRSpec at $1\lesssim z\lesssim9$, which makes it an accessible predictor of the ionization parameter across much of cosmic time. 

Previous works have suggested that ionization parameter-sensitive line ratios such as O32 evolve with redshift \citep[e.g.,][]{Nakajima2014,Sanders2023}. An increasing ionization parameter in star-forming galaxies may cause them to be misidentified as AGN in commonly-used diagnostics \citep[e.g.,][]{Kewley2019b,Cleri2025}, and may cause additional systematics in derived quantities from spectral energy distribution (SED) fitting \citep[e.g.,][]{Byler2017,Johnson2021,Leistedt2023}. 

Studies of the ionization parameter itself beyond cosmic noon ($z\sim2$) have been limited, primarily by sample size and selection function \citep[e.g.,][]{Sanders2023}. This is in part due to the availability of rest-frame optical ionization parameter-sensitive line ratios at high-$z$ being limited to JWST spectroscopy. Additionally, the conversion between observed line ratios and the ionization parameter is dependent on the photoionization modeling assumptions which are often tailored to low-redshift star-forming galaxies \citep[$z\lesssim2$; e.g.,][]{Strom2018,Papovich2022}. In this work, we use JWST spectroscopy and low-redshift comparison samples with large photoionization model suites to demonstrate the evolution of the ionization parameter and discuss the implications of such on other inferred properties. We leverage PRISM and G395M spectroscopy from the Red Unknowns: Bright Infrared Extragalactic Survey \citep[RUBIES;][]{deGraaff2025b} combined with spectroscopy of lower redshift galaxies ($z\lesssim2$) from ground-based surveys to trace the full evolution of the ionization parameter from $0\lesssim z\lesssim9$.

The remainder of this work is as follows. In Section \ref{sec:data}, we discuss the RUBIES spectroscopy and other data products and introduce the low-redshift comparison samples. In Section \ref{sec:results}, we present the evolution of ionization-sensitive line ratios with redshift, stellar mass, and \hb\ luminosity, and demonstrate the inference of the ionization parameter from O32. In Section \ref{sec:discussion}, we discuss the impact of our results on methodologies for estimating the ionization parameter for high-redshift galaxies. In Section \ref{sec:conclusions}, we summarize our results and discuss the future of this science. Where applicable, we employ a flat $\Lambda$CDM cosmology with $H_0 = 70$ km/s/Mpc and $\Omega_m = 0.3$. Unless otherwise noted, all logarithms are base 10. 

\section{Data}\label{sec:data}

\subsection{RUBIES}
We use spectroscopy from the Red Unknowns: Bright Infrared Extragalactic Survey (RUBIES; ID 4233; PIs: A. de Graaff and G. Brammer) JWST Cycle 2 program. The target priority is inverse to number density in F150W-F444W vs F444W color-magnitude space, sampling a diversity of objects. Our sample constitute mainly the lower priority targets of the RUBIES survey, driven by the emission line signal-to-noise constraints. The RUBIES selection, including the selection of high-priority targets (F150W-F444W $>$ 3 and F444W $<$ 27 at $z_\mathrm{phot} > 7$) is described further in \citealt{deGraaff2025b,Lewis2026}. The full details of the survey design and data reduction are described in the survey paper \citep{deGraaff2025b}. The reduced spectra used in this work come from the DAWN JWST Archive\footnote{\href{https://dawn-cph.github.io/dja/}{https://dawn-cph.github.io/dja/}, spectroscopic products DOI \dataset[10.5281/zenodo.7299500]{10.5281/zenodo.7299500}} \citep[DJA;][]{Heintz2024,deGraaff2024}.

RUBIES has NIRSpec MSA \citep{Jakobsen2022,Ferruit2022} PRISM/CLEAR  ($R\equiv\lambda/\Delta\lambda\sim30-330$) spectroscopy covering $0.6-5.3~\mu\mathrm{m}$ and G395M/F290LP ($R\sim 700-1300$) spectroscopy covering $2.9-5.2~\mu\mathrm{m}$ in the Ultra-deep Survey (UDS) and Extended Groth Strip (EGS) fields. The total exposure time per source is 48 min for each disperser/filter combination.

\subsubsection{RUBIES Sample Selection}\label{sec:data:selection}
We restrict all of the subsequent analysis to include only galaxies with coverage of both $\oiii~\lambda5008$ and $\oii~\lambda\lambda3727,3730$. For our primary sample, we require signal-to-noise ratios (\snr) greater than 3 for both \oii\ and \oiii. We additionally form a sample of 297 galaxies which have $\snr>3$ in $\oiii~\lambda5008$ but are not formally detected (i.e., $\snr<3$) in $\oii~\lambda\lambda3727,3730$, which may indicate very high ionization parameter (see Section \ref{sec:discussion} for discussion).  For maximal completeness, we do not attempt to remove active galactic nuclei (AGN) from our sample. Previous studies have demonstrated that commonly-used strong-line selections \citep[e.g.,][]{Baldwin1981,Veilleux1987} are insufficient for predicting the source(s) of ionization in galaxies at high redshifts \citep[e.g.,][]{Cleri2025,Richardson2025}. Additionally, our modeling framework for inferring the ionization parameter accounts for AGN (See Sections \ref{sec:results} and \ref{sec:discussion} for more details). We do check for ``Little Red Dots'' \citep[e.g.,][]{Matthee2024} using the \cite{Hviding2025} catalog, and find that none of the sources in this sample meet their criteria. These selections result in a primary sample of 137 unique RUBIES galaxies. 

For analyses of the specific star formation rates (see Section \ref{sec:results}), we additionally require $\snr>3$ \hb\ detections. This selection introduces a bias toward more actively star forming galaxies, as the \hb\ luminosity is a probe of the near-instantaneous ($<$ 10 Myr) star formation rate \citep[e.g.,][]{Kennicutt1998a,Kennicutt1998b,Kennicutt2012}. The \hb-selected subsample includes 31 RUBIES galaxies. 

\subsubsection{RUBIES Emission Line Fitting}\label{sec:data:lines}
We fit emission lines using \texttt{LiMe}\footnote{\href{https://lime-stable.readthedocs.io/en/latest/}{https://lime-stable.readthedocs.io/en/latest/}}
 v1.3.0 \citep{Fernandez2024}. To confirm the presence of lines, LiMe employs an iterative $n$th-order polynomial to fit the continuum, followed by intensity thresholding that accounts for the calibration flux uncertainty. Finally, LiMe uses LMFIT \citep{Newville2014} to manage the profile constraints and fitting. We select adjacent continuum bands to properly fit the line continuum level and avoid spectroscopic artifacts. For galaxies with RUBIES observations in both the PRISM and G395M modes, we use the averaged line fluxes and their ratios for the following analyses.

\subsubsection{RUBIES SED Fitting and Derived Quantities}\label{sec:data:sed_fitting}
We use stellar masses and other quantities inferred from spectral energy distribution (SED) modeling of photometry from the Near-Infrared Camera on JWST \citep[NIRCam;][]{Rieke2005,Rieke2023}.  The modeling components are briefly reiterated here. Stellar population properties are jointly inferred using the Prospector inference framework \citep{Leja2019,Johnson2021}. The model consists of 18 free parameters, adopting the MIST stellar isochrones \citep{Choi2016,Dotter2016} and MILES stellar library \citep{Sanchez-Blazquez2006} from FSPS \citep{Conroy2010}. The star formation histories assume a non-parametric form, defined by mass formed in 7 logarithmically-spaced time bins \citep[Prospector-alpha][]{Leja2017}. To optimize the photometric inference over the wide parameter space covered by deep JWST surveys \citep[e.g.,][]{Wang2023}, these fits assume a \cite{Chabrier2003} initial mass function (IMF), a Gaussian mass-metallicity prior, and a dynamic star formation history prior $\mathrm{SFH}(\mstar, z)$ from Prospector-$\beta$. A minimum error floor of 5\% is imposed in all fits to reflect the additional systematic uncertainties. Redshift is allowed to vary within the range of $z_{prism} \pm 0.02$. We use the two-screen dust model estimates to correct emission lines for attenuation following a \cite{Kriek2013} attenuation law. For details on the SED modeling, refer to \cite{Wang2024a,Wang2024b}.

\subsection{Comparison Samples}\label{sec:data:comparison}
\subsubsection{SDSS}\label{sec:data:comparison:sdss}
We use emission line fluxes and inferred stellar masses from the Sloan Digital Sky Survey \citep[SDSS;][]{York2000} MPA-JHU Value-Added Catalog\footnote{\href{https://www.sdss4.org/dr14/spectro/galaxy_mpajhu/}{https://www.sdss4.org/dr14/spectro/galaxy\_mpajhu/}} \citep{Brinchmann2004,Kauffmann2003a,Tremonti2004}. SDSS has $R\sim1800$ spectroscopy in the wavelength range $3900\lesssim\lambda\lesssim9100$ \AA\ \citep{Uomoto1999}. The sample selected from SDSS spans $0.04<z<0.1$, where the low redshift limit is chosen to mitigate fiber losses. Stellar masses were inferred using the methodology presented in \cite{Tremonti2004} assuming a \cite{Kroupa2001} IMF. We select all sources in SDSS with \oii, \hb, \oiii, and \ha\ $\snr > 3$. We correct the emission line fluxes using the Balmer decrement, \ha/\hb, assuming a \cite{Calzetti2000} attenuation curve. The final SDSS sample includes 176836 galaxies. 

\subsubsection{LEGA-C}\label{sec:data:comparison:legac}
We use emission line fluxes and inferred stellar masses from the Large Early Galaxy Astrophysics Census \citep[LEGA-C;][]{vanderWel2016}. LEGA-C has $R\sim2500$ spectroscopy from the European Southern Observatory Very Large Telescope (ESO/VLT) Visible Multi-Object Spectrograph \citep[VIMOS;][]{LeFevre2003} for $\sim$3500 K-band-selected galaxies at $0.6<z<1.0$. We use stellar masses and dust attenuation estimates inferred using Prospector from \cite{Nersesian2025} assuming a \cite{Chabrier2003} IMF. We use the two-screen dust model estimates to correct emission lines for attenuation following a \cite{Kriek2013} attenuation law. We note that these attenuation corrections are larger than the other samples presented in this work, which is a selection effect due to the high stellar mass bias of the LEGA-C sample \citep[see][and Section \ref{sec:discussion} for the implications on the results of this work]{vanderWel2016}. We select all sources in LEGA-C with \oii, \hb, and \oiii\  $\snr > 3$, which yields a sample of 236 galaxies. 

\subsubsection{KBSS}\label{sec:data:comparison:kbss}
We use emission line fluxes and inferred stellar masses from the Keck Baryonic Structure Survey \citep[KBSS;][]{Rudie2012,Steidel2014}. KBSS has $R\sim 3600$ spectra with wavelength coverage in the $H$ (1.465-1.799\micron) and $K$ (1.953-2.39\micron) bands obtained using the Multi-Object Spectrometer for InfraRed Exploration \citep[MOSFIRE;][]{McLean2010,McLean2012} for 251 galaxies at $2.0<z<2.6$. Stellar masses were inferred using the methodology presented in \cite{Reddy2012} and \cite{Steidel2014} assuming a \cite{Salpeter1955} IMF. We correct these stellar masses to a \cite{Chabrier2003} IMF for consistency with the other comparison samples \citep[e.g.,][]{Santini2012}. We select all sources in KBSS with \oii, \hb, \oiii, and \ha\ $\snr > 3$. We correct the emission line fluxes using the Balmer decrement, \ha/\hb, assuming a \cite{Calzetti2000} attenuation curve. The final KBSS sample includes 219 galaxies. 

\subsection{Photoionization Models}\label{sec:data:models}
We use photoionization models presented in \cite{Cleri2025}, which were computed using Cloudy \citep[C23.01;][]{Gunasekera2023}. The \cite{Cleri2025} models include both stellar and accreting black hole ionizing spectra. Stellar ionizing spectra come from the Binary Population and Stellar Synthesis library \citep[BPASS v2.2.1][]{Stanway2018} and vary the stellar metallicity: $10^{-5}$, $10^{-4}$, $10^{-3}$, 0.002, 0.003, 0.004, 0.005, 0.006, 0.008, 0.010, 0.014, 0.020, 0.040, where it is taken that $0.020 = \zsol$ \citep[recent works have suggested that $0.0225 = \zsol$, though we maintain the BPASS convention throughout this work;][]{Magg2022}, age $\log(\mathrm{age/yr}) = 6.0$ to $\log(\mathrm{age/yr}) = 11.0$, and IMF (high mass slope $-2.7\leq\alpha_2\leq-2.0$). Black hole accretion disk ionizing spectra come from \cite{Done2012,Cann2018}, and vary over black hole mass ($3\leq\log\mbh/\msun\leq9$). The SEDs can be retrieved from \texttt{XSPEC} \citep{Arnaud1996} using the \textsc{optxagnf} command. The nebular conditions of each model are also varied: ionization parameter is varied from $-4\leq \log U\leq-1$, gas-phase metallicity is matched to the BPASS metallicity grid, and hydrogen density $2\leq \log n_H/cm^{-3}\leq4$. The full description and public release of these models is presented in  \cite{Cleri2025}.

\subsection{Comparison to Simulations}\label{sec:data:comparison:simulations}

We use the publicly available catalogs from the \sphinx\ release \citep{Katz2023c,Rosdahl2018,Rosdahl2022}. The \sphinx\ cosmological simulations model both the large-scale reionization and the ISM processed with detailed radiative transfer for 1380 galaxies in a $20^3$ comoving Mpc volume, with simulated observations for ten sight lines per galaxy. \sphinx\ assumes a \cite{Kroupa2001}-like IMF for BPASS stellar populations.  \sphinx\ provides measurements for 52 nebular emission lines, including all of those used in this work. We select the intrinsic (un-attenuated) emission lines for the following analysis. The \sphinx\ release presents all basic quantities at seven different redshift snapshots: $z \in \{10, 9, 8, 7, 6, 5, 4.64\}$. After $z=4.64$ the volume is completely reionized. The \sphinx\ release is limited to galaxies with star formation rates greater than $0.3\msun~\mathrm{yr}^{-1}$. The quantities derived from the \sphinx\ emission lines and ratios are presented with uncertainties representing their inner 68 percentiles at each redshift. We restrict the \sphinx\ catalog to galaxies with $\log\mstar/\msun>8$ for comparison with the observations and other simulations.

We also compare to star-cLUster-based Model
for Emission liNes \citep[LUMEN;][]{Scharre2026}, which models the emission from star clusters in cosmological simulations to predict emission line fluxes consistent with individual HII regions and integrated galaxy populations. LUMEN assumes a \cite{Chabrier2003} IMF. We present predictions from LUMEN applied to IllustrisTNG50 \citep{Pillepich2018a,Pillepich2019} for simulated galaxies from $3<z<8$. The quantities derived from the LUMEN emission lines and ratios are presented with uncertainties representing their inner 68th percentiles at each redshift. 

LUMEN computes photoionization models with Cloudy v13.03 \citep{Ferland2013} using stellar population synthesis models from \cite{Bruzual2003} and BPASS \citep{Eldridge2017} assuming a \cite{Chabrier2003} IMF from 0.1-100 \msun. LUMEN fixes the hydrogen density of ionized gas to $\mathrm{n_H} = 10^2~\mathrm{cm^{-3}}$. LUMEN includes a prescription for a narrow-line region contribution following \cite{Hirschmann2023}.

The simulated galaxies have systematically lower stellar masses and higher \hb\ luminosities than the observed galaxies. We note that the full \sphinx\ catalog probes down to $\sim2$ dex lower stellar stellar masses than LUMEN or the observed RUBIES galaxies, and the high stellar mass cutoffs are likely due to the small box size of the simulation \citep[see Section \ref{sec:data:comparison:simulations};][]{Rosdahl2018,Rosdahl2022,Katz2023c}. To create a sample more suitable for comparison to the observations, we limit the \sphinx\ catalog to only those galaxies for $\log\mstar/\msun>8$ for the results presented in Section \ref{sec:results}.

\section{Results}\label{sec:results}

\begin{figure*}[ht]%[ht!]
\centering
\epsscale{1.15}
\plotone{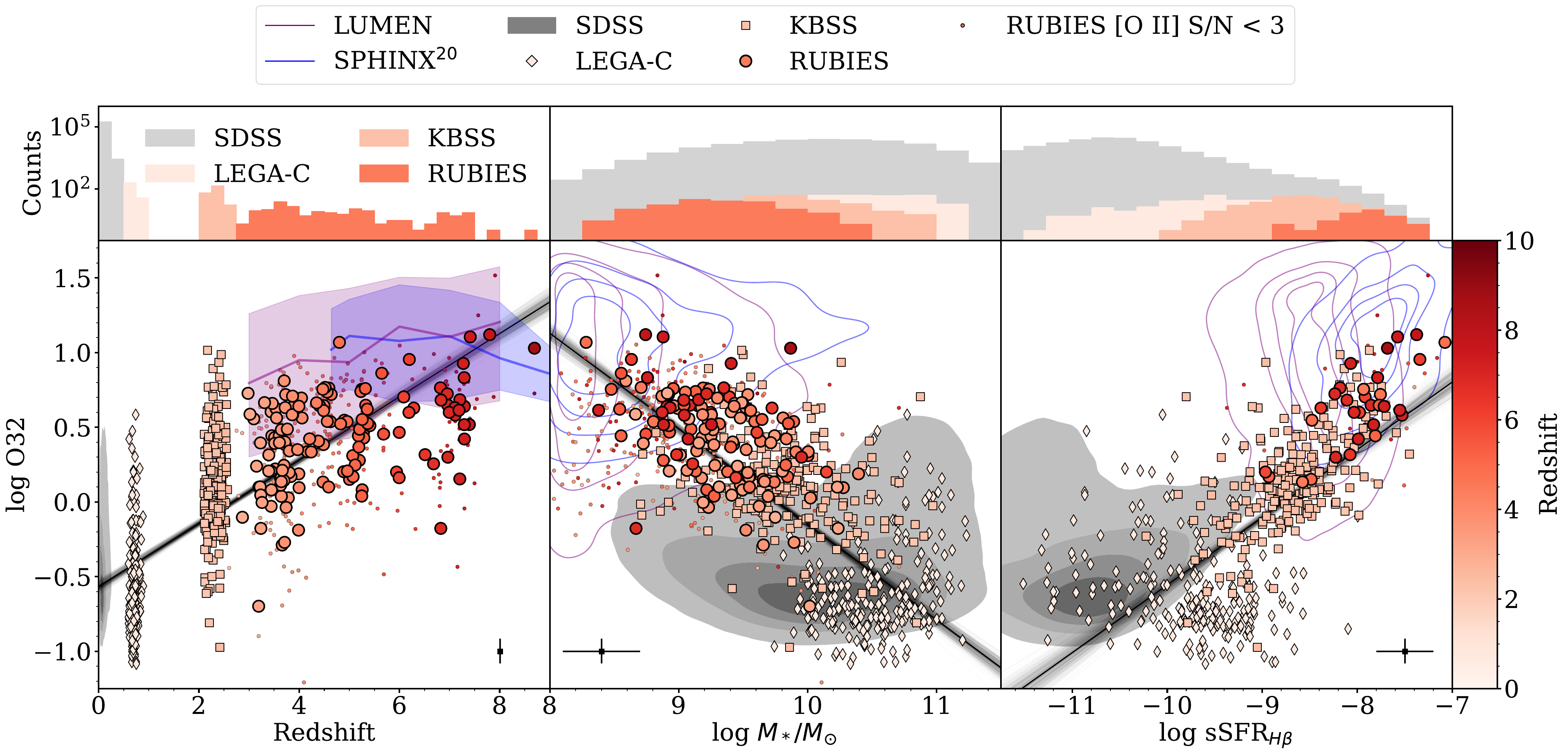}
\caption{The $\log$ O32 ratio (defined in Equation \ref{eq:O32}) as a function of redshift (left), stellar mass $[\msun]$ (center), and sSFR $[\mathrm{yr}^{-1}]$ derived from \hb\ luminosity (right) for galaxies from RUBIES (circles; small circles indicate the \oii\ $\snr<3$ sample), LEGA-C (diamonds), KBSS (hexagons), SDSS (gray contours), and \sphinx\ and LUMEN simulations (blue and purple lines).  The \sphinx\ and LUMEN lines in the left panel include shaded regions bounded by the 16th and 84th percentiles of their distributions in each redshift bin. Non-SDSS points are color coded by redshift. The black lines in each panel show the median linear fit to the observations with gray lines showing 100 draws from the MCMC. The black points in the bottom of each panel show the median uncertainties. The top panels show histograms of the redshift (left), stellar mass (center), and sSFR (right) distributions for each of the observed samples. The fits and summary statistics are reported in Table \ref{tab:fits}.
\label{fig:o32_redshift_mstar_LHb}} 
\end{figure*} 

\begin{figure*}[ht]%[ht!]
\centering
\epsscale{1.15}
\plotone{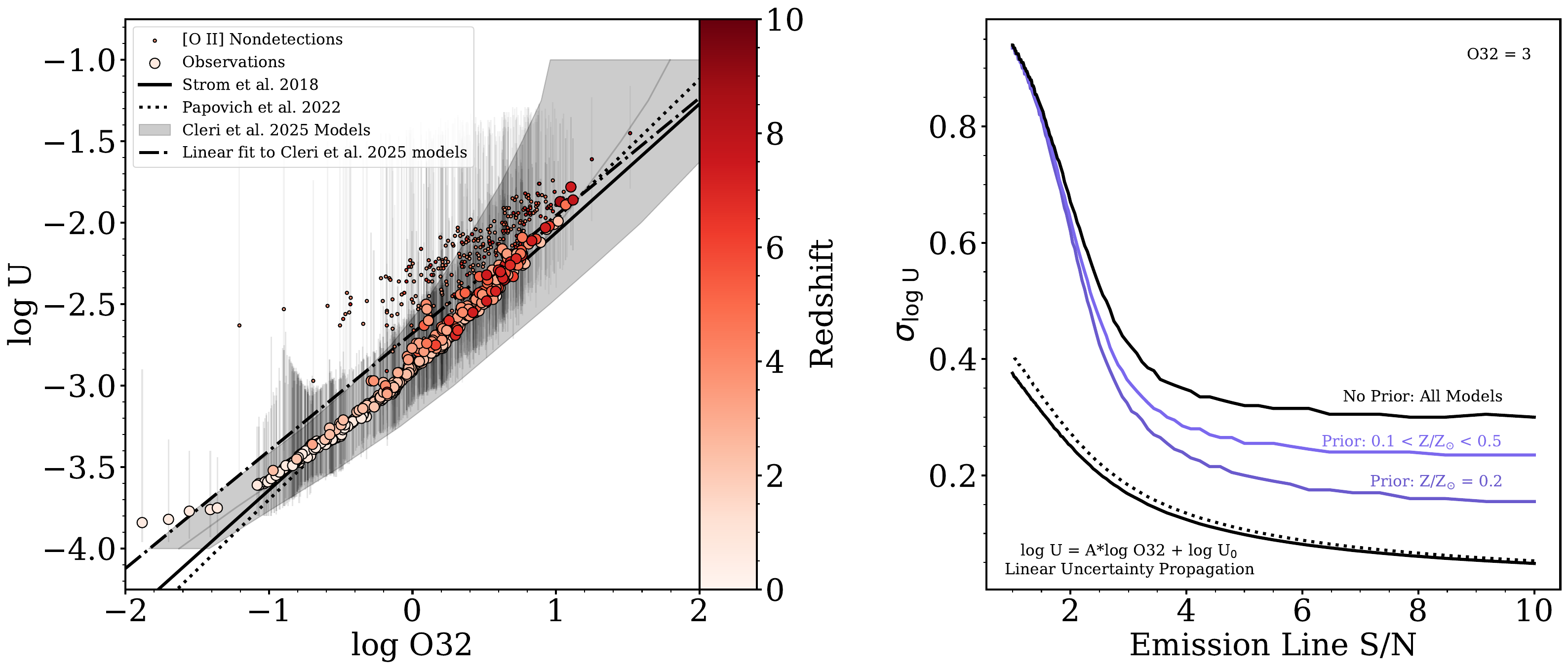}
\caption{Dimensionless ionization parameter $\log$ U as a function of  $\log$ O32, demonstrating the difference in inferred $\log$ U values and uncertainties between the photoionization model inference method presented here and linear best-fit calibrations. We show RUBIES, KBSS, LEGA-C observations as points color-coded by redshift. Small points correspond to the RUBIES \oii\ $\snr<3$ sample. The medians in each $\log$ U step of the \cite{Cleri2025} model grids are shown by the gray line, along with the span of the 16th and 84th percentiles shown as the gray shaded region. Overplotted lines show the \cite{Strom2018} (solid) and \cite{Papovich2022} (dotted) calibrations, along with a linear fit to the \cite{Cleri2025} models (dash-dot). The right panel shows the relation between the uncertainty in $\log$ U determined from the methods in the left panel and the signal to noise ratios of both the \oiii\ and \oii\ emission lines for a fixed O32 = 3. We show three models for the inferred $\log$ U uncertainties, one with an uninformative (flat) prior allowing the full model space on all model parameters (black), and the other with an example prior limiting the model space to $0.1<Z/Z_\odot<0.5$ (light purple), and another with with a prior $Z/Z_\odot=0.2$ (purple). The remaining lines show the uncertainty in $\log$ U as propagated through the linear best-fit calibrations shown in the left panel. 
\label{fig:logU_logO32}} 
\end{figure*} 

\begin{figure*}[ht]%[ht!]
\centering
\epsscale{1.15}
\plotone{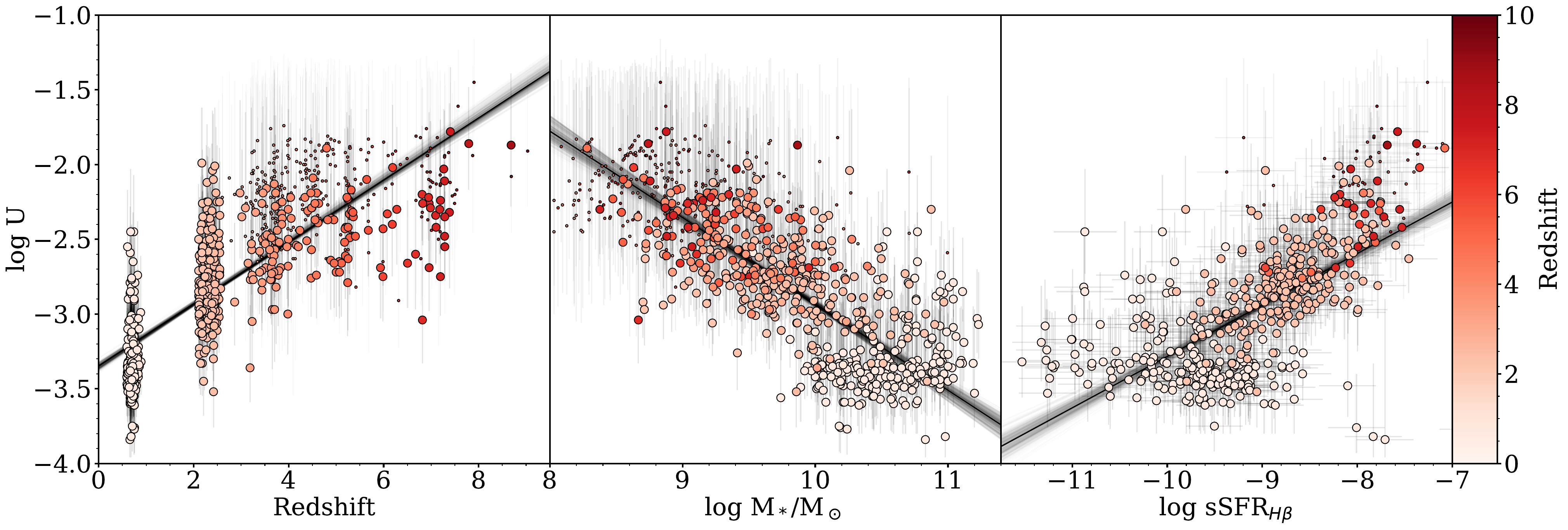}
\caption{Inferred ionization parameter as a function of redshift (left), stellar mass $[\msun]$ (center), and sSFR $[\mathrm{yr}^{-1}]$ derived from \hb\ luminosity (right) for the LEGA-C, KBSS, and RUBIES observations. Small points correspond to the RUBIES \oii\ $\snr<3$ sample. Points are color coded by redshift. The black lines in each panel show the median linear fit with gray lines showing 100 draws from the MCMC. The fits for each panel are reported in Table \ref{tab:fits}.
\label{fig:logU_z_mstar_LHb}} 
\end{figure*} 

\begin{figure*}[ht]%[ht!]
\centering
\epsscale{1.15}
\plotone{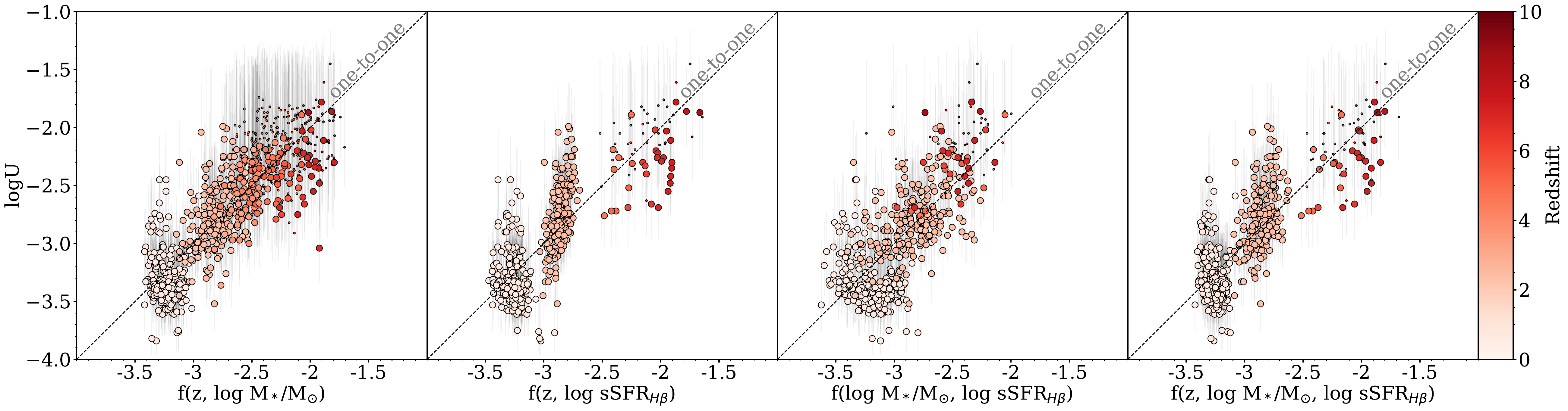}
\caption{Inferred ionization parameter for LEGA-C, KBSS, and RUBIES observations against predictive multivariable fits to other observables and derived quantities (left-to-right): redshift and stellar mass, redshift and sSFR, stellar mass and sSFR, and all three of redshift, stellar mass, and sSFR. Smaller points show the RUBIES subsample of \oii\  $\snr<3$ galaxies. Observations are color-coded by spectroscopic redshift. The dashed black line shows the one-to-one relation. These fits and relevant statistics are reported in Table \ref{tab:fits}.
\label{fig:logU_multivariable_fits}} 
\end{figure*} 

\begin{deluxetable*}{rllllll}[t]
% \digitalasset
\tablecaption{Ionization parameter best-fit models and summary statistics}
\tablehead{
\colhead{Variables} & \colhead{Equation} & \colhead{$\chi^2_\nu$} & \colhead{$\sigma_{int}$} & \colhead{$\sigma_{obs}$}
}
\startdata
$z$ & $\log U = 0.21z -3.35$ & 0.63 & 0.31 & 0.45 \\
$\log \frac{M_*}{M_\odot}$ & $\log U = -0.58\log \frac{M_*}{M_\odot} + 2.85$ & 0.80 & 0.33 & 0.45 \\
$\log sSFR_{\hb}$ & $\log U = 0.34\log sSFR_{\hb} + 0.16$ & 1.97 & 0.47 & 0.34\\
$z$, $\log \frac{M_*}{M_\odot}$ & $\log U = 0.129z-0.278\log \frac{M_*}{M_\odot} - 0.386$ & 0.55 & 0.27 & 0.45 \\
$z$, $\log sSFR_{\hb}$ & $\log U = 0.169z + 0.117\log sSFR_{\hb} - 2.235$ & 0.69 & 0.26 & 0.34 \\
$\log \frac{M_*}{M_\odot}$, $\log sSFR_{\hb}$ & $\log U = -0.393\log \frac{M_*}{M_\odot} + 0.102\log sSFR_{\hb} + 1.920$  & 1.01 & 0.32 & 0.34 \\
$z$, $\log \frac{M_*}{M_\odot}$, $\log sSFR_{\hb}$ & $\log U = 0.153z - 0.172\log \frac{M_*}{M_\odot} + 0.028\log sSFR_{\hb} - 1.28$ & 0.67 & 0.26 & 0.34 
\enddata
\tablecomments{The univariate fits are shown in Figure \ref{fig:logU_z_mstar_LHb}, and the multivariate fits are shown in Figure \ref{fig:logU_multivariable_fits}. We define $\chi^2_\nu$, $\sigma_{int}$, and $\sigma_{obs}$ in Equations \ref{eq:chi_squared_reduced}, \ref{eq:sigma_obs}, and \ref{eq:sigma_meas}, respectively. The units on $sSFR_{\hb}$ are $\mathrm{yr}^{-1}$. We discuss the implications of the fits and their summary statistics in Section \ref{sec:discussion}.}
\label{tab:fits}
\end{deluxetable*}

Here we present the evolution of the ionization-parameter-sensitive O32 ratio from $0\lesssim z\lesssim9$, and infer the dimensionless ionization parameter $\log$ U from photoionization models. We discuss the implications of these analyses and results in Section \ref{sec:discussion}. 

First, we study the direct observable most often used as a tracer of the ionization parameter: O32. Figure \ref{fig:o32_redshift_mstar_LHb} shows the observed O32 ratios from RUBIES, KBSS, LEGA-C, and SDSS against redshift, stellar mass, and specific star formation rate (sSFR $\equiv$ SFR/$\mstar$) derived from \hb\ luminosities, where $\log SFR_{\hb}~\mathrm{[\msun/yr]}= \log L_{\hb}~\mathrm{[erg/s]} - 40.8$ \citep{Kennicutt2012}. Here we require $\snr>3$ \hb\ detections in addition to the O32 lines. This biases the \hb-selected sample to higher inferred SFRs and sSFRs. We compare these observations to simulated galaxies from \sphinx\ and LUMEN, which occupy primarily the low stellar mass and high specific star formation rate parameter space and exhibit high O32 ratios. We find significant correlations between O32 and each of redshift, stellar mass, and \hb\ luminosity, which we fit using \texttt{linmix} \citep{Kelly2007}, a python implementation of an IDL Bayesian model for fitting two-dimensional data with uncertainties in both quantities. The fits to all panels include the RUBIES $\snr<3$ sample to include the potential for very high ionization parameter systems. These results are presented in Table \ref{tab:fits}. The statistics presented in Table \ref{tab:fits} include the following:
\begin{align}\label{eq:chi_squared_reduced}
    \chi^2_\nu &= \frac{1}{d.o.f.}\sum_i \frac{\mathrm{(data_i - model)^2}}{\mathrm{uncertainty_i}^2} 
\end{align}
where $d.o.f.$ is the number of degrees of freedom in the fit, i.e., the number of observations - the number of fit parameters,
\begin{align}\label{eq:sigma_obs}
    \sigma_{int} &= \left(\frac{1}{N}\sum_i \mathrm{(data_i - model)^2} - \mathrm{uncertainty_i}^2\right)^\frac{1}{2} 
\end{align}
\begin{align}\label{eq:sigma_meas}
    \sigma_{obs} &= \left(\frac{1}{N}\sum_i \mathrm{uncertainty_i}^2\right)^\frac{1}{2}
\end{align}
where $N$ is the number of observations.

Next, we infer the ionization parameter from the observed \oiii\ and \oii. Figure \ref{fig:logU_logO32} shows the dimensionless ionization parameter, $\log$ U, against $\log$O32 from the \cite{Cleri2025} photoionization models. We infer ionization parameters from input O32 ratios by calculating likelihoods ($\ln \mathcal{L} \propto \chi^2$) for each model in the \cite{Cleri2025} model library. The priors for the model parameters are determined by the \cite{Cleri2025} model grid, which are uniformly distributed in the dimensionless ionization parameter from $-4\leq\log U\leq -1$. We calculate the 16th, 50th, and 84th percentiles through a linear interpolation between the log U bins of the posteriors. We compare the $\log$ U versus $\log$ O32 relationship in Figure \ref{fig:logU_logO32} with the empirical calibrations from \cite{Strom2018} and \cite{Papovich2022}, along with a linear fit to the \cite{Cleri2025} models. 

We also compare the uncertainties in $\log$ U inferred through the above method to those propagated through linear calibrations \citep[e.g.,][]{Strom2018,Papovich2022}.  We show that the ionization parameter inferred from the full model library leads to larger uncertainties with an uncertainty floor at $\sigma_{\log U}\sim0.3$ for a fixed O32$=$3. The \cite{Strom2018} and \cite{Papovich2022} linear best-fit relations report an average value for log U at a given O32, thus the propagated uncertainties represent the uncertainties in the population mean. The inference methodology presented here marginalizes over the \cite{Cleri2025} model library which has a large spread in log U at a fixed logO32, thus producing a larger and more accurate uncertainty for individual sources.

We explore the impact of informative priors on the inferred uncertainties by taking slices of the model grids (restricted flat priors) when inferring the ionization parameter from \oiii\ and \oii\ fluxes. We show that the uncertainty floor on the inferred ionization parameter is lower than the unrestricted model space, but is still greater than the uncertainties propagated from the \cite{Strom2018} and \cite{Papovich2022} relations. This result indicates that \oiii\ and \oii\ fluxes alone are insufficient to constrain $\log$ U to better than $\sim$0.3 dex for uninformative priors on the \cite{Cleri2025} model grid.

Finally, we connect the ionization parameter to galaxy physical parameters which can be inferred without spectroscopy of \oiii\ and \oii. Figure \ref{fig:logU_z_mstar_LHb} shows the dimensionless ionization parameter, $\log$ U, inferred from the \cite{Cleri2025} photoionization models with the O32 lines as inputs as shown in Figure \ref{fig:logU_logO32}, against redshift, stellar mass, and sSFR for the RUBIES, KBSS, and LEGA-C samples. We report the fits in each panel and their summary statistics in Table \ref{tab:fits}.

Figure \ref{fig:logU_multivariable_fits} compares the dimensionless ionization parameter, $\log$ U, inferred from the \cite{Cleri2025} photoionization models, against multidimensional fits to redshift, stellar mass, and sSFR. These fits are linear in each variable, of the form
\begin{align}
    \log U &= c_0 + \sum_i c_ix_i
\end{align}
where $x_i$ are each subset of the variables $z$, $\log \mstar/\msun$, and $\log L_{\hb}$. These multidimensional fits were calculated using the \texttt{optimize.curvefit} routine from \texttt{scipy} \citep{Virtanen2020} with $\log$ U and the respective combinations of redshift, stellar mass, and \hb\ luminosity as inputs. We show the summary statistics of the fits ($\chi^2_\nu$, $\sigma^2$, and $\sigma^2_{int}$) for each panel, and discuss the implications of these fits in Section \ref{sec:discussion}.

\section{Discussion}\label{sec:discussion}

\subsection{The Redshift Evolution of the Ionization Parameter}\label{sec:discussion:redshift_evolution}

Throughout this work, we have used photoionization models to infer the ionization parameter from observed rest-frame optical emission lines (\oiii\ and \oii) for galaxies from $0<z<9$, with a focus on the $3\lesssim z\lesssim 9$ sample from RUBIES. We show that the observed O32 ratios and inferred log U increase with increasing redshift and specific star formation rate, and decrease with increasing stellar mass (see Figures \ref{fig:o32_redshift_mstar_LHb} and \ref{fig:logU_z_mstar_LHb}). This represents the first time that these relations have been quantified with such a sample at high redshift. We discuss the implications of these results here. 

Previous works have shown that the ionization parameter increases with increasing redshift to at least cosmic noon \citep[e.g.,][]{Nakajima2014,Strom2018,Papovich2022}. In this work, we have shown that this trend continues to $z\sim9$. While redshift is the immediate observable obtained from spectroscopy, we show that the evolution of the ionization parameter with redshift can be more accurately attributed to evolution with redshift and other galaxy properties (i.e. stellar mass and specific star formation rate).

Empirical studies of low-redshift galaxies suggest an anti-correlation between the ionization parameter and gas-phase metallicity \citep[e.g.,][]{Nakajima2014,Steidel2016,Carton2017,Strom2017,Strom2018,Papovich2022}. This is analogous to the correlation between gas-phase metallicity and stellar mass (the ``mass-metallicity relation''), which has been shown to evolve with redshift \citep[e.g.,][]{Tremonti2004,Papovich2022,Laseter2024,Sanders2024,Sanders2026,Lewis2026}. In Figures \ref{fig:o32_redshift_mstar_LHb} and \ref{fig:logU_z_mstar_LHb}, the relationship between the observed O32 and log U with the stellar mass is likely driven by the mass-metallicity relation, where lower metallicity stellar populations emit harder ionizing continua with greater ionizing photon production efficiency \citep[e.g.,][]{Strom2018,Stanway2018,Papovich2022}.

Previous works at low redshifts ($z\lesssim2$) have shown that the ionization parameter increases with increasing specific star formation rate \citep[e.g.,][]{Kaasinen2018}. We find the same behavior continues to higher redshifts, both from O32 and the inferred ionization parameter (see Figure \ref{fig:o32_redshift_mstar_LHb} and \ref{fig:logU_z_mstar_LHb}). We derive star formation rates from \hb\ line luminosities, which trace the H-ionizing radiation ($>13.6$ eV) from young, massive stars. Thus, Balmer line luminosities can be interpreted as a tracer of the near-instantaneous ($<10$Myr) star formation rate \citep[e.g.,][]{Kennicutt1998a,Kennicutt2012}. The correlation between the ionization parameter and the specific star formation rate can be in part interpreted as consistent with the ionizing photon production from H II regions undergoing active star formation, as the ionization parameter is linearly proportional to the local ionizing photon flux \citep[e.g.,][]{Kewley2019b}. Additionally, the ionization parameter is shown to increase with compactness and star formation rate surface density, which have been shown to increase with redshift and specific star formation rate \citep[e.g.,][]{Shibuya2015,Ono2013,Ono2023,Reddy2023b,Ono2025,Yang2025}.

\subsubsection{Toy Model for Redshift Evolution}\label{sec:discussion:toy_model}
Here we construct a toy model to test the physical interpretation of the evolution of the ionization parameter with redshift as shown in Figure \ref{fig:logU_z_mstar_LHb}. We return to the definition of $U$ from e.g., \cite{Kewley2019b}; for an ionizing source in the center of a spherical nebula, 

\begin{align}
    U &= \frac{Q_H}{4 \pi R^2 n_H c}
\end{align}
where $Q_H$ is the H-ionizing photon production, $R$ is the radius of the nebula, and $n_H$ is the hydrogen density. We must first make the simplifying assumption to model a galaxy as a single spherical nebula that satisfies the Str\"omgren condition \citep[$R_S = (3Q_H/4\pi \alpha n_H^2)^{1/3}$;][]{Stromgren1939}, where $\alpha$ is the recombination coefficient and $R_S$ is the Str\"omgren radius. Eliminating the Str\"omgren radius from the definition of the ionization parameter yields the proportionality 
\begin{align}
    U &\propto Q_H^{1/3} n_H^{1/3} \alpha^{-2/3}
\end{align}
which in log space yields
\begin{align}
    \log U &\propto \frac{1}{3}\log Q_H + \frac{1}{3}\log n_H - \frac{2}{3}\log\alpha
\end{align}
We now argue empirically that the density term, $\log U \propto \frac{1}{3}\log n_H$, dominates the redshift evolution. We apply the assumption that the size evolution of galaxies evolves as $R_e\propto (1+z)^{-\beta}$, and that the ISM density scales as the mean volumetric density (i.e., $n_H\propto R_e^{-3}$). We note that this is likely an oversimplifying assumption, and the full diversity of the galaxy size evolution may not be able to be characterized by a simple $R_e\propto (1+z)^{-\beta}$ relation \citep[e.g.,][]{Miller2026}, and that different galaxy types evolve with different slopes \citep[e.g.,][]{Tudorache2026}.  We now have
\begin{align}
    \frac{\partial \log n_H}{\partial \log (1+z)} \approx 3\beta
\end{align}
which further simplifies the density term to 
\begin{align}
    \log U &\propto \beta\log (1+z)
\end{align}
We fit the observations with this nonlinear form and find $\beta = 1.11 \pm 0.03$, which is consistent (within $1-2\sigma$) with empirical size evolution estimates from pre-JWST \citep{Shibuya2015}, and JWST \citep{Yang2025} studies, which found $\beta = 1.10 \pm 0.06$ and $\beta = 1.21 \pm 0.05$, respectively. These results are also consistent with empirical electron density evolution estimates from \cite{Abdurrouf2024}, which found $\beta = 1.2 \pm 0.4$. While there is likely physics captured by the redshift evolution of the ionization parameter in Figure \ref{fig:logU_z_mstar_LHb} due to other sources \citep[e.g., the evolving mass-metallicity and fundamental metallicity relations;][]{Curti2024,Lewis2026}, we attribute the dominant cause to size and density evolution. Future population-level studies of the evolution of electron densities will help constrain the mechanisms driving the increased ionization parameter at high redshifts (see Section \ref{sec:discussion:improvements} for more discussion).

\subsubsection{Comparison to Simulations}\label{sec:discussion:simulations}
In Figure \ref{fig:o32_redshift_mstar_LHb}, we compare observations from RUBIES and our lower redshift comparison samples to predictions for simulated galaxies from \sphinx\ and LUMEN.  At fixed redshifts, \sphinx\ and LUMEN show higher O32, lower stellar mass, and higher sSFRs than the RUBIES observations, though the simulated samples show large scatter. This result is in part due to the redshift range of the \sphinx\ simulations ($4.64\leq z\leq10$) being higher than that of RUBIES ($3\lesssim z\lesssim9$, with a median $z=4.47$), and the $M_{UV}$ selection biasing \sphinx\ toward star-forming galaxies. LUMEN is similarly biased to highly star forming systems by construction, where the Cloudy models are focused on the HII regions around star clusters in IllustrisTNG50 galaxies. These biases may indicate higher ionizing photon production efficiencies or denser nebulae in the simulated galaxies than the observed samples  \citep[e.g.,][]{Rosdahl2018,Rosdahl2022,Katz2023c,Scharre2026}. 

The modeling choices have significant impacts on the results of the simulations. A key limitation of \sphinx\ is the lack of black hole formation or AGN feedback, which is likely important to fully characterize the ionizing conditions and baryon cycling at high redshifts \citep[e.g.,][]{Greene2024,Kocevski2023,Taylor2025a,Taylor2025b,Hviding2025,Scholtz2025}.  \sphinx\ also treats cases where the Str\"omgren sphere is unresolved around multiple star particles in a single cell as a single ionizing source in the center of the cell. This biases the ionization parameter higher than if it were modeled as separate H II regions, leading to elevated ionization parameter-sensitive line ratios (e.g., O32). Additionally, due to the comoving spatial resolution, the resolution of the ISM is worse at lower redshifts in \sphinx; we note that the median of the RUBIES redshifts ($z=4.47$) is near the lowest redshift bin of \sphinx\ ($z=4.64$). We refer the reader to Section 4.3 of \cite{Katz2023c} for further details on the caveats and limitations of the \sphinx\ catalog.

The LUMEN framework makes many of the traditional assumptions for photoionization modeling with Cloudy: idealized one dimensional spherical H II regions with a uniform radiation field incident on a nebula of fixed density. These assumptions are inherently more limited than the multi-component arbitrary geometries capture by other simulations (including \sphinx) or photoionization modeling inference frameworks \citep[e.g.,][see Section \ref{sec:discussion:improvements} for more]{Li2025,Moreschini2026}.  We refer the reader to Section 6 of \cite{Scharre2026} for further details on the caveats and limitations of the LUMEN framework and its applications to the TNG50 simulations.

\subsection{Methodology for Determining the Ionization Parameter from Observations}\label{sec:discussion:logU_methodology}
\subsubsection{Towards More Accurate Uncertainties in Nebular Inference}
Previous works present linear best-fit calibrations to estimate the ionization parameter from observed O32 ratios \citep[e.g.,][]{Strom2018,Papovich2022}. The ionization parameters and associated uncertainties estimated through this method represent those of the mean of the respective samples from \cite{Strom2018} and \cite{Papovich2022}. Throughout this work, we have instead used photoionization models to infer the ionization parameter from the \oiii\ and \oii\ emission line fluxes and their uncertainties for individual galaxies. Here we argue that the inference methodology presented in this work produces more realistic log U uncertainties for individual objects than the linear best-fit calibrations. 

The ionization parameters computed in \cite{Strom2018} are based on a photoionization model grid from \cite{Steidel2016} and \cite{Strom2017} using Cloudy v13.02 \cite{Ferland2013}. \cite{Strom2018} uses BPASS v2 \citep{Stanway2016,Eldridge2017} input ionizing spectra with fixed age \citep[100 Myr;][]{Reddy2012}, varying metallicity $Z/Z_\odot\approx0.07-1.0$,  assuming plane-parallel geometry and a fixed hydrogen density $n_H = 300$ cm$^{-3}$, consistent with the population of $z\sim2.3$ galaxies in KBSS. The full details of their photoionization models can be found in \cite{Steidel2016} and \cite{Strom2017}. 

The ionization parameters computed in \cite{Papovich2022} are calculated using the code ``Inferring the gas-phase metallicity (Z) and Ionization parameter'' \citep[IZI;][]{Blanc2015}. \cite{Papovich2022} computes posteriors for ionization parameter using MAPPINGS-V \citep{Kewley2019a} assuming fixed pressure over a grid of gas phase metallicities and ionization parameters. The full details of their modeling and inference of ionization parameters can be found in Section 4.2 of \cite{Papovich2022}. 

We show in Figure \ref{fig:logU_logO32} that the uncertainties derived from propagating through a linear calibration are systematically smaller than those estimated by the inference framework described in this work. The ionization parameter is often degenerate with single emission line ratios (e.g., O32). In our methodology, we capture the full range of degeneracies by using the broad suite of models from \cite{Cleri2025}, thus accurately modeling the true uncertainty as verified by mock testing. We do note that these large uncertainties result in sub-unity $\chi^2_\nu$ (See Table \ref{tab:fits}), which may indicate that the observed sample does not span the range of physical parameters probed by the \cite{Cleri2025} models. This can be accounted for by including informative priors, e.g., constraining the gas-phase metallicity (right panel of \ref{fig:logU_logO32}).  We demonstrate that the inclusion of additional constraints decreases the inferred uncertainties and lowers the uncertainty ``floor'', meaning that there still exists a minimum uncertainty at arbitrarily small measurement uncertainties. In future work, we will explore the inclusion of more emission lines to reduce our prior dependence (see Section \ref{sec:discussion:improvements} for discussion).   

To complement the above analyses, we also present calibrations to calculate ionization parameter from observations and derived quantities other than the O32 ratio. We show the relations between ionization parameter and redshift, stellar mass, and specific star formation rate in Figure \ref{fig:logU_z_mstar_LHb}, and the multivariable fits to combinations of these three parameters in Figure \ref{fig:logU_multivariable_fits}. This represents a metric to determine the ionization parameter, with large scatter, for galaxies without spectroscopy available. Importantly, the redshift dependence suggests galaxies do evolve with time, beyond just their mass and sSFR. To illustrate this result, consider three galaxies each with LMC-like stellar mass and specific star formation rate \citep[$\log\mstar/\msun \approx3\times10^9$ and $\log$ sSFR$/\mathrm{yr}^{-1}\approx -10.8$;][]{Tacchella2022} at redshifts $z=0$, $z=2$, and $z=6$; the relations in Table \ref{tab:fits} predict their ionization parameters to be log U $=$ -3.3 for the galaxy at $z=0$, log U $=$ -2.9 for the galaxy at $z=2$, and log U $=$ -2.3 for the galaxy at $z=6$. This result suggests that the physics driving the ionization parameter evolution with cosmic time is not fully encapsulated by the stellar mass and specific star formation rate, though each of these are independently informative.

% For example, galaxies with LMC-like stellar mass and specific star formation rate ($\log\mstar/\msun \approx3\times10^9$ and $\log$ sSFR $\approx -10.8$, \citet{Tacchella2022}) will have higher ionization parameters at higher redshifts according to the relation shown in Table \ref{tab:fits} (-3.3 at $z=0$, -2.9 at $z=2$, and -2.3 at $z=6$). This result suggests that the physics driving the ionization parameter evolution with cosmic time is not fully encapsulated by the stellar mass and specific star formation rate, though each of these are independently informative. 

As an added benefit, this analysis invites a method to determine the ionization parameter without spectroscopy of specific emission lines. This result is useful for applications such as SED fitting, where it is often necessary to assume a single ionization parameter for all galaxies, which immediately introduces degeneracies \citep[e.g.,][]{Byler2017,Johnson2021,Leistedt2023}. Conveniently, the required parameters from Figure \ref{fig:logU_multivariable_fits} and the relevant fits in Table \ref{tab:fits} are standard inferred quantities of SED fitting routines \citep[e.g.,][]{Leja2017,Leja2019,Johnson2021}.

\subsubsection{Caveats and Improvements for Future Work}\label{sec:discussion:improvements}
The methodology of determining ionization parameters can be improved beyond the simplifications made in this work. One such improvement is the allowance for multiple clouds/ionization zones \citep[e.g., Highly Optimised Emission-line Ratios Using photoioNisation (\texttt{HOMERUN});][]{Marconi2024}. Recent works have used \texttt{HOMERUN} to demonstrate that the model of a single cloud ionized by a single source is not sufficient to reproduce the observed spectra in three galaxies at $2<z<6$ \citep[][]{Moreschini2026}. \cite{Moreschini2026} show that the ionization parameter from an individual spectrum can vary by as much as $\sim$0.4 dex when allowing for unresolved high density clouds ($n_{e,max} = 10^7$ cm$^{-3}$), and that these high density clouds can account for nearly half of the $\oiii~\lambda5008$ emission.

Another improvement to this methodology is the incorporation of more emission lines and other spectral features. Contemporary methodology such as \texttt{HOMERUN}, \texttt{MULTIGRIS} \citep{Lebouteiller2022}, and \texttt{Cue} \citep{Li2025} simultaneously model multiple emission lines, thus leveraging more information than single strong line ratio diagnostics. The inclusion of additional spectral features in the inference can simultaneously inform the ionization parameter and other nebular conditions (e.g., gas-phase metallicity, hydrogen density) and the ionizing continuum (e.g., stellar populations versus black hole accretion disks, initial mass function and other stellar population properties) to allow for a more complete interpretation of observations. This problem is a ripe use case for modern statistical methods including simulation-based inference \citep[e.g.,][]{Cranmer2020,Deistler2025}. 

We look forward to upcoming spectroscopy to include this additional data in future works studying the nebular conditions of high-redshift galaxies. Particularly, the collisional deexcitation of \oii\ confounds the derivation of log U from O32 with the higher electron densities observed in higher-$z$ systems \citep[e.g.,][]{Isobe2023,Abdurrouf2024,Fujimoto2024a,Harikane2025,Topping2025,Choustikov2026,Rogers2026,Scharre2026}. This is due to the different critical densities of the \oii\ and \oiii\ transitions: $n_{c,\lambda3727}\sim1.5\times 10^4~\mathrm{cm^{-3}}$ and $n_{c,\lambda3729}\sim3.4\times 10^3~\mathrm{cm^{-3}}$, $n_{c,\lambda5008}\sim6.8\times 10^5~\mathrm{cm^{-3}}$ \citep[all assuming $T=10^4$K;][]{Osterbrock1989,Osterbrock2006}. This represents a key limitation of estimating the ionization parameter from O32, where log U and other ISM properties can only be constrained with simultaneous information about the electron densities. These constraints can be made with high-resolution ($R\sim2700$) NIRSpec grating spectra to be available in upcoming JWST programs (e.g., the Characterizing Electron Densities at Reionization (CEDAR) survey; program ID: 10518, PI: N. J. Cleri, N. J. Cleri et al. in preparation).

\section{Conclusions}\label{sec:conclusions}
In this work, we study the evolution of the ionization parameter, $\log$ U, with redshift and other observed and derived quantities for 434 galaxies in the RUBIES survey. We compile literature sources from SDSS, LEGA-C, and KBSS to form a sample which spans $0\lesssim z \lesssim 9$. Our primary findings are as follows:

\begin{itemize}
    \item We show that the ionization parameter-sensitive line ratio O32 increases with increasing redshift in Figure \ref{fig:o32_redshift_mstar_LHb}. We also find that O32 decreases with increasing stellar mass and increases with increasing specific star formation rate. 
    \item We infer ionization parameters based on \oiii\ and \oii\ using photoionization models from \cite{Cleri2025}, and find log U medians broadly consistent with linear best-fit calibrations, though with systematically larger uncertainties (see Figure \ref{fig:logU_logO32}). We argue that the larger uncertainties inferred from the photoionization models are more representative of the true uncertainty on an individual log U estimate with O32 as the only information. We show that the introduction of an informative prior on the model grid parameters (e.g., gas-phase metallicity) improves the uncertainties on the inferred ionization parameter, though there exists an uncertainty floor which is an indicator of the incompleteness of information derivable from O32 alone. We discuss the implications of these inferred $\log$ U values and their uncertainties in Section \ref{sec:discussion:logU_methodology}. 
    \item We show that the inferred ionization parameters increase with increasing redshift and specific star formation rate, and decrease with increasing stellar mass. We generalize this result to three and four dimensions by providing multidimensional fits to predict $\log$ U from combinations of redshift, stellar mass, and specific star formation rate.  We discuss the implications of these fits and their utility in applications such as  SED fitting where \oiii\ and \oii\ may not always be available through spectroscopy in Section \ref{sec:discussion}.

    \item Our fits show that the physics driving the evolution of the ionization parameter is not fully encapsulated by the stellar mass or sSFR variations alone, thus the redshift encodes additional information. The toy model presented in Section \ref{sec:discussion:toy_model} suggests that the dominant mechanism behind the redshift evolution of the ionization parameter is consistent with the variations in electron density due to size evolution.
\end{itemize}

Future work in this field will require additional information beyond \oiii\ and \oii\ emission lines to characterize the gas conditions in high-redshift galaxies. We discuss in Section \ref{sec:discussion} several methods for adding this information, including modeling the multi-phase ISM and including more emission lines in the inference framework. We look forward to upcoming high-resolution spectroscopy from JWST (e.g., CEDAR) to simultaneously model the temperature, density, and ionization structures in statistical samples of high-redshift galaxies.

\section*{Acknowledgments}
We are deeply thankful to all of those who have dedicated themselves to making JWST a reality. N.J.C. thanks the entire RUBIES team for their support and valuable contributions. N.J.C. also thanks Taylor Hutchison for valuable discussions, and the anonymous referee for their contributions to improving the manuscript. 

N.J.C. acknowledges support from JWST-AR-05558, as well as funding from the Eberly Postdoctoral Fellowship from the Eberly College of Science at the Pennsylvania State University. 

MH, LS, and AP acknowledge funding from the Swiss National Science Foundation (SNF) via a PRIMA Grant PR00P2 193577 “From cosmic dawn to high noon: the role of black holes for young galaxies”.

AdG acknowledges support from a Clay Fellowship awarded by the Smithsonian Astrophysical Observatory.

The computations of the photoionization models from \cite{Cleri2025} which are presented in this work were performed using the Texas A\&M High Performance Research Computing cluster Grace\footnote{See \url{https://hprc.tamu.edu/kb/User-Guides/Grace}, where the namesake of the cluster is  \href{https://en.wikipedia.org/wiki/Grace_Hopper}{Grace Hopper}.}.

Some of the computations for this research were performed on the Pennsylvania State University's Institute for Computational and Data Sciences' Roar supercomputer. This publication made use of the NASA Astrophysical Data System (ADS) and Science Explorer (SciX) for bibliographic information.

The Pennsylvania State University campuses are located on the original homelands of the Erie, Haudenosaunee (Seneca, Cayuga, Onondaga, Oneida, Mohawk, and Tuscarora), Lenape (Delaware Nation, Delaware Tribe, Stockbridge-Munsee), Monongahela, Shawnee (Absentee, Eastern, and Oklahoma), Susquehannock, and Wahzhazhe (Osage) Nations. As a land grant institution, we acknowledge and honor the traditional caretakers of these lands and strive to understand and model their responsible stewardship. We also acknowledge the longer history of these lands and our place in that history.

\bibliography{main}{}
\bibliographystyle{mn2e}

\end{document}

%% file: macros.tex
\newcommand{\vdag}{(v)^\dagger}
\newcommand\aastex{AAS\TeX}
\newcommand\latex{La\TeX}

\newcommand{\Lya}{\hbox{{\rm Ly}$\alpha$}}
\newcommand{\lya}{\hbox{{\rm Ly}$\alpha$}}
% Balmer
\newcommand{\Ha}{\hbox{{\rm H}$\alpha$}}
\newcommand{\ha}{\hbox{{\rm H}$\alpha$}}
\newcommand{\Hb}{\hbox{{\rm H}$\beta$}}
\newcommand{\hb}{\hbox{{\rm H}$\beta$}}
\newcommand{\Hg}{\hbox{{\rm H}$\gamma$}}
\newcommand{\hg}{\hbox{{\rm H}$\gamma$}}
% Paschen
\newcommand{\Paa}{\hbox{{\rm Pa}$\alpha$}}
\newcommand{\paa}{\hbox{{\rm Pa}$\alpha$}}
\newcommand{\Pab}{\hbox{{\rm Pa}$\beta$}}
\newcommand{\pab}{\hbox{{\rm Pa}$\beta$}}

% Other lines by atomic number
% He
\newcommand{\heii}{\hbox{\ion{He}{2}}}
% C
\newcommand{\cii}{\hbox{[\ion{C}{2}]}}
\newcommand{\ciii}{\hbox{[\ion{C}{3}]}}
\newcommand{\civ}{\hbox{\ion{C}{4}}}
% N
\newcommand{\nii}{\hbox{[\ion{N}{2}]}}
\newcommand{\niv}{\hbox{[\ion{N}{4}]}}
\newcommand{\nv}{\hbox{[\ion{N}{5}]}}
% O
\newcommand{\oi}{\hbox{\ion{O}{1}}}
\newcommand{\oii}{\hbox{[\ion{O}{2}]}}
\newcommand{\xoii}{\hbox{[\ion{O}{2}] $\lambda\lambda 3727,3729$}}
\newcommand{\oiii}{\hbox{[\ion{O}{3}]}}
\newcommand{\xoiii}{\hbox{[\ion{O}{3}] $\lambda\lambda 4960,5008$}}
\newcommand{\xoiiia}{\hbox{[\ion{O}{3}]$\lambda 5008$}}
\newcommand{\xoiiib}{\hbox{[\ion{O}{3}]$\lambda 4960$}}
\newcommand{\oiv}{\hbox{[\ion{O}{4}]}}
% Ne
\newcommand{\neii}{\hbox{[\ion{Ne}{2}]}}
\newcommand{\neiii}{\hbox{[\ion{Ne}{3}]}}
\newcommand{\xneiii}{\hbox{[\ion{Ne}{3}] $\lambda 3870$}}
\newcommand{\nev}{\hbox{[\ion{Ne}{5}]}}
\newcommand{\xnev}{\hbox{[\ion{Ne}{5}] $\lambda 3347,3427$}}
\newcommand{\xneva}{\hbox{[\ion{Ne}{5}] $\lambda 3427$}}
% S
\newcommand{\sii}{\hbox{[\ion{S}{2}]}}
% Ar
\newcommand{\ariii}{\hbox{[\ion{Ar}{3}]}}
\newcommand{\ariv}{\hbox{[\ion{Ar}{4}]}}
% Mg
\newcommand{\mgii}{\hbox{\ion{Mg}{2}}}
% Fe
\newcommand{\feii}{\hbox{[\ion{Fe}{2}}]}
\newcommand{\fex}{\hbox{[\ion{Fe}{10}}]}

% Measurements from lines
\newcommand{\vlya}{\hbox{$\Delta$v$_{Ly\alpha}$}}
\newcommand{\vciv}{\hbox{$\Delta$v$_{CIV}$}}
\newcommand{\wciii}{\hbox{W$_{\text{\sc C\,iii}]}$}~}
\newcommand{\wciv}{\hbox{W$_{\text{\sc C\,iv}}$}~}
\newcommand{\wlya}{\hbox{W$_{\text{Ly$\alpha$}}$}~}
\newcommand{\fesc}{\hbox{$f_{esc}$}}

% Telescopes
\newcommand{\hst}{\textit{HST}}
\newcommand{\hstlong}{\textit{Hubble Space Telescope}}
\newcommand{\jwst}{\textit{JWST}}
\newcommand{\jwstlong}{\textit{James Webb Space Telescope}}
\newcommand{\romanlong}{\textit{Nancy Grace Roman Space Telescope}}
\newcommand{\HST}{{\it HST}}
\newcommand{\JWST}{{\it JWST}}
\newcommand{\spitzer}{{\it Spitzer}}
\newcommand{\Spitzer}{{\it Spitzer}}

% objects/targets
\newcommand{\gndone}{z7\_GND\_42912}
\newcommand{\gndtwo}{z7\_GND\_16863}
\newcommand{\gndthree}{z7\_GND\_22483}

% code & models
\newcommand{\cloudy}{{\sc Cloudy}}
\newcommand{\bpass}{{\sc bpass}}
\newcommand{\grizli}{\texttt{grizli}}
\newcommand{\pyneb}{\texttt{PyNeb}}

\newcommand{\snr}{\hbox{S/N}}

\newcommand{\logoh}{\hbox{$\log$(O/H)}}
\newcommand{\sphinx}{\hbox{SPHINX$^{20}$}}
\newcommand{\mstar}{\hbox{M$_*$}}
\newcommand{\msun}{\hbox{M$_\odot$}}
\newcommand{\msol}{\hbox{M$_\odot$}}
\newcommand{\zsun}{\hbox{Z$_\odot$}}
\newcommand{\zsol}{\hbox{Z$_\odot$}}

\newcommand{\mbh}{\hbox{M$_\mathrm{BH}$}}

\newcommand{\Hplus}{\hbox{H$^{+}$}}

\newcommand{\Oplus}{\hbox{O$^{+}$}}
\newcommand{\Otwoplus}{\hbox{O$^{2+}$}}
\newcommand{\Othreeplus}{\hbox{O$^{3+}$}}

\newcommand{\Netwoplus}{\hbox{Ne$^{2+}$}}

\newcommand{\njc}[1]{\textcolor{red}{\textbf{\texttt{#1}}}}

%% file: authors.tex
\author{Nikko J. Cleri$^{1,2,3}$\thanks{$^*$E-mail: \href{mailto:cleri@psu.edu}{cleri@psu.edu}}, 
Zach J. Lewis$^{4}$, 
Joel Leja$^{1,2,3}$,
Jakob M. Helton$^{1}$,
Emilie Burnham$^{1,3}$,
Olivia Curtis$^{1}$,
Anna de Graaff$^{5,6}$,
Michaela Hirschmann$^{7}$,
Harley Katz$^{8}$,
Michael V. Maseda$^{4}$,
Ian McConachie$^{4}$,
Ad\'ele~Plat$^{7}$,
Lucie~Scharr\'e$^{7}$
}

\affiliation{$^{1}$Department of Astronomy and Astrophysics, The Pennsylvania State University, University Park, PA 16802, USA}
\affiliation{$^{2}$Institute for Computational and Data Sciences, The Pennsylvania State University, University Park, PA 16802, USA}
\affiliation{$^{3}$Institute for Gravitation and the Cosmos, The Pennsylvania State University, University Park, PA 16802, USA}
\affiliation{$^{4}$Department of Astronomy, University of Wisconsin-Madison, Madison, WI 53706, USA}
\affiliation{$^{5}$Center for Astrophysics, Harvard \& Smithsonian, 60 Garden St, Cambridge, MA 02138, USA}
\affiliation{$^{6}$Max-Planck-Institut f\"ur Astronomie, K\"onigstuhl 17, D-69117 Heidelberg, Germany}
\affiliation{$^{7}$Institute of Physics, Laboratory for Galaxy Evolution, EPFL, Observatory of Sauverny, Chemin Pegasi 51, CH-1290 Versoix, Switzerland}
\affiliation{$^{8}$Department of Astronomy \& Astrophysics, University of Chicago, Chicago, IL 60637, USA}

%% file: main.bib
@ARTICLE{Abdurrouf2024,
       author = {{Abdurro'uf} and {Larson}, Rebecca L. and {Coe}, Dan and {Hsiao}, Tiger Yu-Yang and {{\'A}lvarez-M{\'a}rquez}, Javier and {G{\'o}mez}, Alejandro Crespo and {Adamo}, Angela and {Bhatawdekar}, Rachana and {Bik}, Arjan and {Bradley}, Larry D. and {Conselice}, Christopher J. and {Dayal}, Pratika and {Diego}, Jose M. and {Fujimoto}, Seiji and {Furtak}, Lukas J. and {Hutchison}, Taylor A. and {Jung}, Intae and {Killi}, Meghana and {Kokorev}, Vasily and {Mingozzi}, Matilde and {Norman}, Colin and {Resseguier}, Tom and {Ricotti}, Massimo and {Rigby}, Jane R. and {Vanzella}, Eros and {Welch}, Brian and {Windhorst}, Rogier A. and {Xu}, Xinfeng and {Zitrin}, Adi},
        title = "{JWST NIRSpec High-resolution Spectroscopy of MACS0647{\textendash}JD at z = 10.167: Resolved [O II] Doublet and Electron Density in an Early Galaxy}",
      journal = {\apj},
         year = 2024,
        month = sep,
       volume = {973},
       number = {1},
          eid = {47},
        pages = {47},
          doi = {10.3847/1538-4357/ad6001},
archivePrefix = {arXiv},
       eprint = {2404.16201},
 primaryClass = {astro-ph.GA},
       adsurl = {https://ui.adsabs.harvard.edu/abs/2024ApJ...973...47A}
}

@INPROCEEDINGS{Arnaud1996,
       author = {{Arnaud}, K.~A.},
        title = "{XSPEC: The First Ten Years}",
    booktitle = {Astronomical Data Analysis Software and Systems V},
         year = 1996,
       editor = {{Jacoby}, George H. and {Barnes}, Jeannette},
       series = {Astronomical Society of the Pacific Conference Series},
       volume = {101},
        month = jan,
        pages = {17},
       adsurl = {https://ui.adsabs.harvard.edu/abs/1996ASPC..101...17A}
}

@ARTICLE{Baldwin1981,
       author = {{Baldwin}, J.~A. and {Phillips}, M.~M. and {Terlevich}, R.},
        title = "{Classification parameters for the emission-line spectra of extragalactic objects.}",
      journal = {\pasp},
         year = 1981,
        month = feb,
       volume = {93},
        pages = {5-19},
          doi = {10.1086/130766},
       adsurl = {https://ui.adsabs.harvard.edu/abs/1981PASP...93....5B}
}

@ARTICLE{Berg2021,
       author = {{Berg}, Danielle A. and {Chisholm}, John and {Erb}, Dawn K. and {Skillman}, Evan D. and {Pogge}, Richard W. and {Olivier}, Grace M.},
        title = "{Characterizing Extreme Emission-line Galaxies. I. A Four-zone Ionization Model for Very High-ionization Emission}",
      journal = {\apj},
         year = 2021,
        month = dec,
       volume = {922},
       number = {2},
          eid = {170},
        pages = {170},
          doi = {10.3847/1538-4357/ac141b},
archivePrefix = {arXiv},
       eprint = {2105.12765},
 primaryClass = {astro-ph.GA},
       adsurl = {https://ui.adsabs.harvard.edu/abs/2021ApJ...922..170B}
}

@ARTICLE{Blanc2015,
       author = {{Blanc}, Guillermo A. and {Kewley}, Lisa and {Vogt}, Fr{\'e}d{\'e}ric P.~A. and {Dopita}, Michael A.},
        title = "{IZI: Inferring the Gas Phase Metallicity (Z) and Ionization Parameter (q) of Ionized Nebulae Using Bayesian Statistics}",
      journal = {\apj},
         year = 2015,
        month = jan,
       volume = {798},
       number = {2},
          eid = {99},
        pages = {99},
          doi = {10.1088/0004-637X/798/2/99},
archivePrefix = {arXiv},
       eprint = {1410.8146},
 primaryClass = {astro-ph.GA},
       adsurl = {https://ui.adsabs.harvard.edu/abs/2015ApJ...798...99B}
}

@ARTICLE{Brinchmann2004,
       author = {{Brinchmann}, J. and {Charlot}, S. and {White}, S.~D.~M. and {Tremonti}, C. and {Kauffmann}, G. and {Heckman}, T. and {Brinkmann}, J.},
        title = "{The physical properties of star-forming galaxies in the low-redshift Universe}",
      journal = {\mnras},
         year = 2004,
        month = jul,
       volume = {351},
       number = {4},
        pages = {1151-1179},
          doi = {10.1111/j.1365-2966.2004.07881.x},
archivePrefix = {arXiv},
       eprint = {astro-ph/0311060},
 primaryClass = {astro-ph},
       adsurl = {https://ui.adsabs.harvard.edu/abs/2004MNRAS.351.1151B}
}

@ARTICLE{Bruzual2003,
   author = {{Bruzual}, G. and {Charlot}, S.},
    title = "{Stellar population synthesis at the resolution of 2003}",
  journal = {\mnras},
   eprint = {astro-ph/0309134},
     year = 2003,
    month = oct,
   volume = 344,
    pages = {1000-1028},
      doi = {10.1046/j.1365-8711.2003.06897.x},
   adsurl = {http://adsabs.harvard.edu/abs/2003MNRAS.344.1000B}
}

@ARTICLE{Byler2017,
       author = {{Byler}, Nell and {Dalcanton}, Julianne J. and {Conroy}, Charlie and {Johnson}, Benjamin D.},
        title = "{Nebular Continuum and Line Emission in Stellar Population Synthesis Models}",
      journal = {\apj},
         year = 2017,
        month = may,
       volume = {840},
       number = {1},
          eid = {44},
        pages = {44},
          doi = {10.3847/1538-4357/aa6c66},
archivePrefix = {arXiv},
       eprint = {1611.08305},
 primaryClass = {astro-ph.GA},
       adsurl = {https://ui.adsabs.harvard.edu/abs/2017ApJ...840...44B}
}

@ARTICLE{Calzetti1994,
       author = {{Calzetti}, Daniela and {Kinney}, Anne L. and {Storchi-Bergmann}, Thaisa},
        title = "{Dust Extinction of the Stellar Continua in Starburst Galaxies: The Ultraviolet and Optical Extinction Law}",
      journal = {\apj},
         year = 1994,
        month = jul,
       volume = {429},
        pages = {582},
          doi = {10.1086/174346},
       adsurl = {https://ui-adsabs-harvard-edu.ezproxy.lib.uconn.edu/abs/1994ApJ...429..582C}
}

@ARTICLE{Calzetti2000,
   author = {{Calzetti}, D. and {Armus}, L. and {Bohlin}, R.~C. and {Kinney}, A.~L. and 
	{Koornneef}, J. and {Storchi-Bergmann}, T.},
    title = "{The Dust Content and Opacity of Actively Star-forming Galaxies}",
  journal = {\apj},
   eprint = {astro-ph/9911459},
     year = 2000,
    month = apr,
   volume = 533,
    pages = {682-695},
      doi = {10.1086/308692},
   adsurl = {http://adsabs.harvard.edu/abs/2000ApJ...533..682C}
}

@ARTICLE{Cann2018,
       author = {{Cann}, Jenna M. and {Satyapal}, Shobita and {Abel}, Nicholas P. and {Ricci}, Claudio and {Secrest}, Nathan J. and {Blecha}, Laura and {Gliozzi}, Mario},
        title = "{The Hunt for Intermediate-mass Black Holes in the JWST Era}",
      journal = {\apj},
         year = 2018,
        month = jul,
       volume = {861},
       number = {2},
          eid = {142},
        pages = {142},
          doi = {10.3847/1538-4357/aac64a},
archivePrefix = {arXiv},
       eprint = {1805.09351},
 primaryClass = {astro-ph.GA},
       adsurl = {https://ui.adsabs.harvard.edu/abs/2018ApJ...861..142C}
}

@ARTICLE{Carton2017,
       author = {{Carton}, David and {Brinchmann}, Jarle and {Shirazi}, Maryam and {Contini}, Thierry and {Epinat}, Beno{\^\i}t and {Erroz-Ferrer}, Santiago and {Marino}, Raffaella A. and {Martinsson}, Thomas P.~K. and {Richard}, Johan and {Patr{\'\i}cio}, Vera},
        title = "{Inferring gas-phase metallicity gradients of galaxies at the seeing limit: a forward modelling approach}",
      journal = {\mnras},
         year = 2017,
        month = jun,
       volume = {468},
       number = {2},
        pages = {2140-2163},
          doi = {10.1093/mnras/stx545},
archivePrefix = {arXiv},
       eprint = {1703.01090},
 primaryClass = {astro-ph.GA},
       adsurl = {https://ui.adsabs.harvard.edu/abs/2017MNRAS.468.2140C}
}

@ARTICLE{Chabrier2003,
       author = {{Chabrier}, Gilles},
        title = "{Galactic Stellar and Substellar Initial Mass Function}",
      journal = {Publications of the Astronomical Society of the Pacific},
         year = 2003,
        month = Jul,
       volume = {115},
        pages = {763-795},
          doi = {10.1086/376392},
       adsurl = {https://ui.adsabs.harvard.edu/#abs/2003PASP..115..763C}
}

@ARTICLE{Choi2016,
       author = {{Choi}, Jieun and {Dotter}, Aaron and {Conroy}, Charlie and {Cantiello}, Matteo and {Paxton}, Bill and {Johnson}, Benjamin D.},
        title = "{Mesa Isochrones and Stellar Tracks (MIST). I. Solar-scaled Models}",
      journal = {\apj},
         year = 2016,
        month = jun,
       volume = {823},
       number = {2},
          eid = {102},
        pages = {102},
          doi = {10.3847/0004-637X/823/2/102},
archivePrefix = {arXiv},
       eprint = {1604.08592},
 primaryClass = {astro-ph.SR},
       adsurl = {https://ui.adsabs.harvard.edu/abs/2016ApJ...823..102C}
}

@ARTICLE{Choustikov2026,
       author = {{Choustikov}, Nicholas and {Katz}, Harley and {Cameron}, Alex J. and {Saxena}, Aayush and {Devriendt}, Julien and {Slyz}, Adrianne and {Rey}, Martin P. and {Cadiou}, Corentin and {Blaizot}, Jeremy and {Kimm}, Taysun and et al.},
        title = "{MEGATRON: Disentangling Physical Processes and Observational Bias in the Multi-Phase ISM of High-Redshift Galaxies}",
      journal = {The Open Journal of Astrophysics},
         year = 2026,
        month = feb,
       volume = {9},
        pages = {58199},
          doi = {10.33232/001c.158199},
archivePrefix = {arXiv},
       eprint = {2510.06347},
 primaryClass = {astro-ph.GA},
       adsurl = {https://ui.adsabs.harvard.edu/abs/2026OJAp....958199C}
}

@ARTICLE{Cleri2025,
       author = {{Cleri}, Nikko J. and {Olivier}, Grace M. and {Backhaus}, Bren E. and {Leja}, Joel and {Papovich}, Casey and {Trump}, Jonathan R. and {Arrabal Haro}, Pablo and {Buat}, V{\'e}ronique and {Burgarella}, Denis and {Burnham}, Emilie and et al.},
        title = "{Optical Strong Line Ratios Cannot Distinguish between Stellar Populations and Accreting Black Holes at High Ionization Parameters and Low Metallicities}",
      journal = {\apj},
         year = 2025,
        month = dec,
       volume = {994},
       number = {2},
          eid = {146},
        pages = {146},
          doi = {10.3847/1538-4357/ae0f17},
archivePrefix = {arXiv},
       eprint = {2506.21660},
 primaryClass = {astro-ph.GA},
       adsurl = {https://ui.adsabs.harvard.edu/abs/2025ApJ...994..146C}
}

@ARTICLE{Conroy2010,
   author = {{Conroy}, C. and {Gunn}, J.~E.},
    title = "{The Propagation of Uncertainties in Stellar Population Synthesis Modeling. III. Model Calibration, Comparison, and Evaluation}",
  journal = {\apj},
archivePrefix = "arXiv",
   eprint = {0911.3151},
     year = 2010,
    month = apr,
   volume = 712,
    pages = {833-857},
      doi = {10.1088/0004-637X/712/2/833},
   adsurl = {http://adsabs.harvard.edu/abs/2010ApJ...712..833C}
}

@ARTICLE{Cranmer2020,
       author = {{Cranmer}, Kyle and {Brehmer}, Johann and {Louppe}, Gilles},
        title = "{The frontier of simulation-based inference}",
      journal = {Proceedings of the National Academy of Science},
         year = 2020,
        month = dec,
       volume = {117},
       number = {48},
        pages = {30055-30062},
          doi = {10.1073/pnas.1912789117},
archivePrefix = {arXiv},
       eprint = {1911.01429},
 primaryClass = {stat.ML},
       adsurl = {https://ui.adsabs.harvard.edu/abs/2020PNAS..11730055C}
}

@ARTICLE{Cullen2021,
       author = {{Cullen}, F. and {Shapley}, A.~E. and {McLure}, R.~J. and {Dunlop}, J.~S. and {Sanders}, R.~L. and {Topping}, M.~W. and {Reddy}, N.~A. and {Amor{\'\i}n}, R. and {Begley}, R. and {Bolzonella}, M. and {Calabr{\`o}}, A. and {Carnall}, A.~C. and {Castellano}, M. and {Cimatti}, A. and {Cirasuolo}, M. and {Cresci}, G. and {Fontana}, A. and {Fontanot}, F. and {Garilli}, B. and {Guaita}, L. and {Hamadouche}, M. and {Hathi}, N.~P. and {Mannucci}, F. and {McLeod}, D.~J. and {Pentericci}, L. and {Saxena}, A. and {Talia}, M. and {Zamorani}, G.},
        title = "{The NIRVANDELS Survey: a robust detection of {\ensuremath{\alpha}}-enhancement in star-forming galaxies at z ≃ 3.4}",
      journal = {\mnras},
         year = 2021,
        month = jul,
       volume = {505},
       number = {1},
        pages = {903-920},
          doi = {10.1093/mnras/stab1340},
archivePrefix = {arXiv},
       eprint = {2103.06300},
 primaryClass = {astro-ph.GA},
       adsurl = {https://ui.adsabs.harvard.edu/abs/2021MNRAS.505..903C}
}

@ARTICLE{Curti2024,
       author = {{Curti}, Mirko and {Maiolino}, Roberto and {Curtis-Lake}, Emma and {Chevallard}, Jacopo and {Carniani}, Stefano and {D'Eugenio}, Francesco and {Looser}, Tobias J. and {Scholtz}, Jan and {Charlot}, Stephane and {Cameron}, Alex and et al.},
        title = "{JADES: Insights into the low-mass end of the mass-metallicity-SFR relation at 3 < z < 10 from deep JWST/NIRSpec spectroscopy}",
      journal = {\aap},
         year = 2024,
        month = apr,
       volume = {684},
          eid = {A75},
        pages = {A75},
          doi = {10.1051/0004-6361/202346698},
archivePrefix = {arXiv},
       eprint = {2304.08516},
 primaryClass = {astro-ph.GA},
       adsurl = {https://ui.adsabs.harvard.edu/abs/2024A&A...684A..75C}
}

@ARTICLE{deGraaff2024,
       author = {{de Graaff}, Anna and {Rix}, Hans-Walter and {Carniani}, Stefano and {Suess}, Katherine A. and {Charlot}, St{\'e}phane and {Curtis-Lake}, Emma and {Arribas}, Santiago and {Baker}, William M. and {Boyett}, Kristan and {Bunker}, Andrew J. and {Cameron}, Alex J. and {Chevallard}, Jacopo and {Curti}, Mirko and {Eisenstein}, Daniel J. and {Franx}, Marijn and {Hainline}, Kevin and {Hausen}, Ryan and {Ji}, Zhiyuan and {Johnson}, Benjamin D. and {Jones}, Gareth C. and {Maiolino}, Roberto and {Maseda}, Michael V. and {Nelson}, Erica and {Parlanti}, Eleonora and {Rawle}, Tim and {Robertson}, Brant and {Tacchella}, Sandro and {{\"U}bler}, Hannah and {Williams}, Christina C. and {Willmer}, Christopher N.~A. and {Willott}, Chris},
        title = "{Ionised gas kinematics and dynamical masses of z {\ensuremath{\gtrsim}} 6 galaxies from JADES/NIRSpec high-resolution spectroscopy}",
      journal = {\aap},
         year = 2024,
        month = apr,
       volume = {684},
          eid = {A87},
        pages = {A87},
          doi = {10.1051/0004-6361/202347755},
archivePrefix = {arXiv},
       eprint = {2308.09742},
 primaryClass = {astro-ph.GA},
       adsurl = {https://ui.adsabs.harvard.edu/abs/2024A&A...684A..87D}
}

@ARTICLE{deGraaff2025b,
       author = {{de Graaff}, Anna and {Brammer}, Gabriel and {Weibel}, Andrea and {Lewis}, Zach and {Maseda}, Michael V. and {Oesch}, Pascal A. and {Bezanson}, Rachel and {Boogaard}, Leindert A. and {Cleri}, Nikko J. and {Cooper}, Olivia R. and {Gottumukkala}, Rashmi and {Greene}, Jenny E. and {Hirschmann}, Michaela and {Hviding}, Raphael E. and {Katz}, Harley and {Labb{\'e}}, Ivo and {Leja}, Joel and {Matthee}, Jorryt and {McConachie}, Ian and {Miller}, Tim B. and {Naidu}, Rohan P. and {Price}, Sedona H. and {Rix}, Hans-Walter and {Setton}, David J. and {Suess}, Katherine A. and {Wang}, Bingjie and {Whitaker}, Katherine E. and {Williams}, Christina C.},
        title = "{RUBIES: A complete census of the bright and red distant Universe with JWST/NIRSpec}",
      journal = {\aap},
         year = 2025,
        month = may,
       volume = {697},
          eid = {A189},
        pages = {A189},
          doi = {10.1051/0004-6361/202452186},
archivePrefix = {arXiv},
       eprint = {2409.05948},
 primaryClass = {astro-ph.GA},
       adsurl = {https://ui.adsabs.harvard.edu/abs/2025A&A...697A.189D}
}

@ARTICLE{Deistler2025,
       author = {{Deistler}, Michael and {Boelts}, Jan and {Steinbach}, Peter and {Moss}, Guy and {Moreau}, Thomas and {Gloeckler}, Manuel and {Rodrigues}, Pedro L.~C. and {Linhart}, Julia and {Lappalainen}, Janne K. and {Miller}, Benjamin Kurt and {Gon{\c{c}}alves}, Pedro J. and {Lueckmann}, Jan-Matthis and {Schr{\"o}der}, Cornelius and {Macke}, Jakob H.},
        title = "{Simulation-Based Inference: A Practical Guide}",
      journal = {arXiv e-prints},
         year = 2025,
        month = aug,
          eid = {arXiv:2508.12939},
        pages = {arXiv:2508.12939},
          doi = {10.48550/arXiv.2508.12939},
archivePrefix = {arXiv},
       eprint = {2508.12939},
 primaryClass = {stat.ML},
       adsurl = {https://ui.adsabs.harvard.edu/abs/2025arXiv250812939D}
}

@ARTICLE{Done2012,
       author = {{Done}, Chris and {Davis}, S.~W. and {Jin}, C. and {Blaes}, O. and {Ward}, M.},
        title = "{Intrinsic disc emission and the soft X-ray excess in active galactic nuclei}",
      journal = {\mnras},
         year = 2012,
        month = mar,
       volume = {420},
       number = {3},
        pages = {1848-1860},
          doi = {10.1111/j.1365-2966.2011.19779.x},
archivePrefix = {arXiv},
       eprint = {1107.5429},
 primaryClass = {astro-ph.HE},
       adsurl = {https://ui.adsabs.harvard.edu/abs/2012MNRAS.420.1848D}
}

@ARTICLE{Dotter2016,
       author = {{Dotter}, Aaron},
        title = "{MESA Isochrones and Stellar Tracks (MIST) 0: Methods for the Construction of Stellar Isochrones}",
      journal = {\apjs},
         year = 2016,
        month = jan,
       volume = {222},
       number = {1},
          eid = {8},
        pages = {8},
          doi = {10.3847/0067-0049/222/1/8},
archivePrefix = {arXiv},
       eprint = {1601.05144},
 primaryClass = {astro-ph.SR},
       adsurl = {https://ui.adsabs.harvard.edu/abs/2016ApJS..222....8D}
}

@ARTICLE{Eldridge2017,
       author = {{Eldridge}, J.~J. and {Stanway}, E.~R. and {Xiao}, L. and {McClelland}, L.~A.~S. and {Taylor}, G. and {Ng}, M. and {Greis}, S.~M.~L. and {Bray}, J.~C.},
        title = "{Binary Population and Spectral Synthesis Version 2.1: Construction, Observational Verification, and New Results}",
      journal = {\pasa},
         year = 2017,
        month = nov,
       volume = {34},
          eid = {e058},
        pages = {e058},
          doi = {10.1017/pasa.2017.51},
archivePrefix = {arXiv},
       eprint = {1710.02154},
 primaryClass = {astro-ph.SR},
       adsurl = {https://ui.adsabs.harvard.edu/abs/2017PASA...34...58E}
}

@ARTICLE{Erb2006,
       author = {{Erb}, Dawn K. and {Shapley}, Alice E. and {Pettini}, Max and {Steidel},
        Charles C. and {Reddy}, Naveen A. and {Adelberger}, Kurt L.},
        title = "{The Mass-Metallicity Relation at z>~2}",
      journal = {\apj},
         year = 2006,
        month = Jun,
       volume = {644},
        pages = {813-828},
          doi = {10.1086/503623},
       adsurl = {https://ui.adsabs.harvard.edu/#abs/2006ApJ...644..813E}
}

@ARTICLE{Ferland2013,
       author = {{Ferland}, G.~J. and {Porter}, R.~L. and {van Hoof}, P.~A.~M. and {Williams}, R.~J.~R. and {Abel}, N.~P. and {Lykins}, M.~L. and {Shaw}, G. and {Henney}, W.~J. and {Stancil}, P.~C.},
        title = "{The 2013 Release of Cloudy}",
      journal = {\rmxaa},
         year = 2013,
        month = apr,
       volume = {49},
        pages = {137-163},
          doi = {10.48550/arXiv.1302.4485},
archivePrefix = {arXiv},
       eprint = {1302.4485},
 primaryClass = {astro-ph.GA},
       adsurl = {https://ui.adsabs.harvard.edu/abs/2013RMxAA..49..137F}
}

@ARTICLE{Fernandez2024,
       author = {{Fern{\'a}ndez}, V. and {Amor{\'\i}n}, R. and {Firpo}, V. and {Morisset}, C.},
        title = "{LIME: A LIne MEasuring library for large and complex spectroscopic data sets. I. Implementation of a virtual observatory for JWST spectra}",
      journal = {\aap},
         year = 2024,
        month = aug,
       volume = {688},
          eid = {A69},
        pages = {A69},
          doi = {10.1051/0004-6361/202449224},
archivePrefix = {arXiv},
       eprint = {2405.15072},
 primaryClass = {astro-ph.IM},
       adsurl = {https://ui.adsabs.harvard.edu/abs/2024A&A...688A..69F}
}

@ARTICLE{Ferruit2022,
       author = {{Ferruit}, P. and {Jakobsen}, P. and {Giardino}, G. and {Rawle}, T. and {Alves de Oliveira}, C. and {Arribas}, S. and {Beck}, T.~L. and {Birkmann}, S. and {B{\"o}ker}, T. and {Bunker}, A.~J. and et al.},
        title = "{The Near-Infrared Spectrograph (NIRSpec) on the James Webb Space Telescope. II. Multi-object spectroscopy (MOS)}",
      journal = {\aap},
         year = 2022,
        month = may,
       volume = {661},
          eid = {A81},
        pages = {A81},
          doi = {10.1051/0004-6361/202142673},
archivePrefix = {arXiv},
       eprint = {2202.03306},
 primaryClass = {astro-ph.IM},
       adsurl = {https://ui.adsabs.harvard.edu/abs/2022A&A...661A..81F}
}

@ARTICLE{Fujimoto2024a,
       author = {{Fujimoto}, Seiji and {Ouchi}, Masami and {Nakajima}, Kimihiko and {Harikane}, Yuichi and {Isobe}, Yuki and {Brammer}, Gabriel and {Oguri}, Masamune and {Gim{\'e}nez-Arteaga}, Clara and {Heintz}, Kasper E. and {Kokorev}, Vasily and {Bauer}, Franz E. and {Ferrara}, Andrea and {Kojima}, Takashi and {Lagos}, Claudia del P. and {Laura}, Sommovigo and {Schaerer}, Daniel and {Shimasaku}, Kazuhiro and {Hatsukade}, Bunyo and {Kohno}, Kotaro and {Sun}, Fengwu and {Valentino}, Francesco and {Watson}, Darach and {Fudamoto}, Yoshinobu and {Inoue}, Akio K. and {Gonz{\'a}lez-L{\'o}pez}, Jorge and {Koekemoer}, Anton M. and {Knudsen}, Kirsten and {Lee}, Minju M. and {Magdis}, Georgios E. and {Richard}, Johan and {Strait}, Victoria B. and {Sugahara}, Yuma and {Tamura}, Yoichi and {Toft}, Sune and {Umehata}, Hideki and {Walth}, Gregory},
        title = "{JWST and ALMA Multiple-line Study in and around a Galaxy at z = 8.496: Optical to Far-Infrared Line Ratios and the Onset of an Outflow Promoting Ionizing Photon Escape}",
      journal = {\apj},
         year = 2024,
        month = apr,
       volume = {964},
       number = {2},
          eid = {146},
        pages = {146},
          doi = {10.3847/1538-4357/ad235c},
archivePrefix = {arXiv},
       eprint = {2212.06863},
 primaryClass = {astro-ph.GA},
       adsurl = {https://ui.adsabs.harvard.edu/abs/2024ApJ...964..146F}
}

@ARTICLE{Greene2024,
       author = {{Greene}, Jenny E. and {Labbe}, Ivo and {Goulding}, Andy D. and {Furtak}, Lukas J. and {Chemerynska}, Iryna and {Kokorev}, Vasily and {Dayal}, Pratika and {Volonteri}, Marta and {Williams}, Christina C. and {Wang}, Bingjie and {Setton}, David J. and {Burgasser}, Adam J. and {Bezanson}, Rachel and {Atek}, Hakim and {Brammer}, Gabriel and {Cutler}, Sam E. and {Feldmann}, Robert and {Fujimoto}, Seiji and {Glazebrook}, Karl and {de Graaff}, Anna and {Khullar}, Gourav and {Leja}, Joel and {Marchesini}, Danilo and {Maseda}, Michael V. and {Matthee}, Jorryt and {Miller}, Tim B. and {Naidu}, Rohan P. and {Nanayakkara}, Themiya and {Oesch}, Pascal A. and {Pan}, Richard and {Papovich}, Casey and {Price}, Sedona H. and {van Dokkum}, Pieter and {Weaver}, John R. and {Whitaker}, Katherine E. and {Zitrin}, Adi},
        title = "{UNCOVER Spectroscopy Confirms the Surprising Ubiquity of Active Galactic Nuclei in Red Sources at z > 5}",
      journal = {\apj},
         year = 2024,
        month = mar,
       volume = {964},
       number = {1},
          eid = {39},
        pages = {39},
          doi = {10.3847/1538-4357/ad1e5f},
archivePrefix = {arXiv},
       eprint = {2309.05714},
 primaryClass = {astro-ph.GA},
       adsurl = {https://ui.adsabs.harvard.edu/abs/2024ApJ...964...39G}
}

@ARTICLE{Gunasekera2023,
       author = {{Gunasekera}, Chamani M. and {van Hoof}, Peter A.~M. and {Chatzikos}, Marios and {Ferland}, Gary J.},
        title = "{The 23.01 Release of Cloudy}",
      journal = {Research Notes of the American Astronomical Society},
         year = 2023,
        month = nov,
       volume = {7},
       number = {11},
          eid = {246},
        pages = {246},
          doi = {10.3847/2515-5172/ad0e75},
archivePrefix = {arXiv},
       eprint = {2311.10163},
 primaryClass = {astro-ph.GA},
       adsurl = {https://ui.adsabs.harvard.edu/abs/2023RNAAS...7..246G}
}

@ARTICLE{Harikane2025,
       author = {{Harikane}, Yuichi and {Sanders}, Ryan L. and {Ellis}, Richard and {Jones}, Tucker and {Ouchi}, Masami and {Laporte}, Nicolas and {Roberts-Borsani}, Guido and {Katz}, Harley and {Nakajima}, Kimihiko and {Ono}, Yoshiaki and {Gupta}, Mansi},
        title = "{JWST and ALMA Joint Analysis with [O II] {\ensuremath{\lambda}}{\ensuremath{\lambda}}3726, 3729, [O III] {\ensuremath{\lambda}}4363, [O III] 88 {\ensuremath{\mu}}m, and [O III] 52 {\ensuremath{\mu}}m: Multizone Evolution of Electron Densities at z {\ensuremath{\sim}} 0─14 and its Impact on Metallicity Measurements}",
      journal = {\apj},
         year = 2025,
        month = nov,
       volume = {993},
       number = {2},
          eid = {204},
        pages = {204},
          doi = {10.3847/1538-4357/ae0e53},
archivePrefix = {arXiv},
       eprint = {2505.09186},
 primaryClass = {astro-ph.GA},
       adsurl = {https://ui.adsabs.harvard.edu/abs/2025ApJ...993..204H}
}

@ARTICLE{Heintz2024,
       author = {{Heintz}, Kasper E. and {Watson}, Darach and {Brammer}, Gabriel and {Vejlgaard}, Simone and {Hutter}, Anne and {Strait}, Victoria B. and {Matthee}, Jorryt and {Oesch}, Pascal A. and {Jakobsson}, P{\'a}ll and {Tanvir}, Nial R. and {Laursen}, Peter and {Naidu}, Rohan P. and {Mason}, Charlotte A. and {Killi}, Meghana and {Jung}, Intae and {Hsiao}, Tiger Yu-Yang and {Abdurro'uf} and {Coe}, Dan and {Arrabal Haro}, Pablo and {Finkelstein}, Steven L. and {Toft}, Sune},
        title = "{Strong damped Lyman-{\ensuremath{\alpha}} absorption in young star-forming galaxies at redshifts 9 to 11}",
      journal = {Science},
         year = 2024,
        month = may,
       volume = {384},
       number = {6698},
        pages = {890-894},
          doi = {10.1126/science.adj0343},
archivePrefix = {arXiv},
       eprint = {2306.00647},
 primaryClass = {astro-ph.GA},
       adsurl = {https://ui.adsabs.harvard.edu/abs/2024Sci...384..890H}
}

@ARTICLE{Hirschmann2023,
       author = {{Hirschmann}, Michaela and {Charlot}, Stephane and {Feltre}, Anna and {Curtis-Lake}, Emma and {Somerville}, Rachel S. and {Chevallard}, Jacopo and {Choi}, Ena and {Nelson}, Dylan and {Morisset}, Christophe and {Plat}, Adele and {Vidal-Garcia}, Alba},
        title = "{Emission-line properties of IllustrisTNG galaxies: from local diagnostic diagrams to high-redshift predictions for JWST}",
      journal = {\mnras},
         year = 2023,
        month = dec,
       volume = {526},
       number = {3},
        pages = {3610-3636},
          doi = {10.1093/mnras/stad2955},
archivePrefix = {arXiv},
       eprint = {2212.02522},
 primaryClass = {astro-ph.GA},
       adsurl = {https://ui.adsabs.harvard.edu/abs/2023MNRAS.526.3610H}
}

@ARTICLE{Hviding2025,
       author = {{Hviding}, Raphael E. and {de Graaff}, Anna and {Miller}, Tim B. and {Setton}, David J. and {Greene}, Jenny E. and {Labb{\'e}}, Ivo and {Brammer}, Gabriel and {Bezanson}, Rachel and {Boogaard}, Leindert A. and {Cleri}, Nikko J. and et al.},
        title = "{RUBIES: A spectroscopic census of little red dots: All point sources with v-shaped continua have broad lines}",
      journal = {\aap},
         year = 2025,
        month = oct,
       volume = {702},
          eid = {A57},
        pages = {A57},
          doi = {10.1051/0004-6361/202555816},
       adsurl = {https://ui.adsabs.harvard.edu/abs/2025A&A...702A..57H}
}

@ARTICLE{Isobe2023,
       author = {{Isobe}, Yuki and {Ouchi}, Masami and {Nakajima}, Kimihiko and {Harikane}, Yuichi and {Ono}, Yoshiaki and {Xu}, Yi and {Zhang}, Yechi and {Umeda}, Hiroya},
        title = "{Redshift Evolution of Electron Density in the Interstellar Medium at z   0-9 Uncovered with JWST/NIRSpec Spectra and Line-spread Function Determinations}",
      journal = {\apj},
         year = 2023,
        month = oct,
       volume = {956},
       number = {2},
          eid = {139},
        pages = {139},
          doi = {10.3847/1538-4357/acf376},
archivePrefix = {arXiv},
       eprint = {2301.06811},
 primaryClass = {astro-ph.GA},
       adsurl = {https://ui.adsabs.harvard.edu/abs/2023ApJ...956..139I}
}

@ARTICLE{Jakobsen2022,
       author = {{Jakobsen}, P. and {Ferruit}, P. and {Alves de Oliveira}, C. and {Arribas}, S. and {Bagnasco}, G. and {Barho}, R. and {Beck}, T.~L. and {Birkmann}, S. and {B{\"o}ker}, T. and {Bunker}, A.~J. and {Charlot}, S. and {de Jong}, P. and {de Marchi}, G. and {Ehrenwinkler}, R. and {Falcolini}, M. and {Fels}, R. and {Franx}, M. and {Franz}, D. and {Funke}, M. and {Giardino}, G. and {Gnata}, X. and {Holota}, W. and {Honnen}, K. and {Jensen}, P.~L. and {Jentsch}, M. and {Johnson}, T. and {Jollet}, D. and {Karl}, H. and {Kling}, G. and {K{\"o}hler}, J. and {Kolm}, M. -G. and {Kumari}, N. and {Lander}, M.~E. and {Lemke}, R. and {L{\'o}pez-Caniego}, M. and {L{\"u}tzgendorf}, N. and {Maiolino}, R. and {Manjavacas}, E. and {Marston}, A. and {Maschmann}, M. and {Maurer}, R. and {Messerschmidt}, B. and {Moseley}, S.~H. and {Mosner}, P. and {Mott}, D.~B. and {Muzerolle}, J. and {Pirzkal}, N. and {Pittet}, J. -F. and {Plitzke}, A. and {Posselt}, W. and {Rapp}, B. and {Rauscher}, B.~J. and {Rawle}, T. and {Rix}, H. -W. and {R{\"o}del}, A. and {Rumler}, P. and {Sabbi}, E. and {Salvignol}, J. -C. and {Schmid}, T. and {Sirianni}, M. and {Smith}, C. and {Strada}, P. and {te Plate}, M. and {Valenti}, J. and {Wettemann}, T. and {Wiehe}, T. and {Wiesmayer}, M. and {Willott}, C.~J. and {Wright}, R. and {Zeidler}, P. and {Zincke}, C.},
        title = "{The Near-Infrared Spectrograph (NIRSpec) on the James Webb Space Telescope. I. Overview of the instrument and its capabilities}",
      journal = {\aap},
         year = 2022,
        month = may,
       volume = {661},
          eid = {A80},
        pages = {A80},
          doi = {10.1051/0004-6361/202142663},
archivePrefix = {arXiv},
       eprint = {2202.03305},
 primaryClass = {astro-ph.IM},
       adsurl = {https://ui.adsabs.harvard.edu/abs/2022A&A...661A..80J}
}

@ARTICLE{Johnson2021,
       author = {{Johnson}, Benjamin D. and {Leja}, Joel and {Conroy}, Charlie and {Speagle}, Joshua S.},
        title = "{Stellar Population Inference with Prospector}",
      journal = {\apjs},
         year = 2021,
        month = jun,
       volume = {254},
       number = {2},
          eid = {22},
        pages = {22},
          doi = {10.3847/1538-4365/abef67},
archivePrefix = {arXiv},
       eprint = {2012.01426},
 primaryClass = {astro-ph.GA},
       adsurl = {https://ui.adsabs.harvard.edu/abs/2021ApJS..254...22J}
}

@ARTICLE{Kaasinen2018,
       author = {{Kaasinen}, Melanie and {Kewley}, Lisa and {Bian}, Fuyan and {Groves}, Brent and {Kashino}, Daichi and {Silverman}, John and {Kartaltepe}, Jeyhan},
        title = "{The ionization parameter of star-forming galaxies evolves with the specific star formation rate}",
      journal = {\mnras},
         year = 2018,
        month = jul,
       volume = {477},
       number = {4},
        pages = {5568-5589},
          doi = {10.1093/mnras/sty1012},
archivePrefix = {arXiv},
       eprint = {1804.10621},
 primaryClass = {astro-ph.GA},
       adsurl = {https://ui.adsabs.harvard.edu/abs/2018MNRAS.477.5568K}
}

@ARTICLE{Katz2023c,
       author = {{Katz}, Harley and {Rosdahl}, Joki and {Kimm}, Taysun and {Blaizot}, Jeremy and {Choustikov}, Nicholas and {Farcy}, Marion and {Garel}, Thibault and {Haehnelt}, Martin G. and {Michel-Dansac}, Leo and {Ocvirk}, Pierre},
        title = "{The SPHINX Public Data Release: Forward Modelling High-Redshift JWST Observations with Cosmological Radiation Hydrodynamics Simulations}",
      journal = {The Open Journal of Astrophysics},
         year = 2023,
        month = dec,
       volume = {6},
          eid = {44},
        pages = {44},
          doi = {10.21105/astro.2309.03269},
archivePrefix = {arXiv},
       eprint = {2309.03269},
 primaryClass = {astro-ph.GA},
       adsurl = {https://ui.adsabs.harvard.edu/abs/2023OJAp....6E..44K}
}

@ARTICLE{Kauffmann2003a,
       author = {{Kauffmann}, Guinevere and {Heckman}, Timothy M. and {White}, Simon D.~M. and {Charlot}, St{\'e}phane and {Tremonti}, Christy and {Brinchmann}, Jarle and {Bruzual}, Gustavo and {Peng}, Eric W. and {Seibert}, Mark and {Bernardi}, Mariangela and {Blanton}, Michael and {Brinkmann}, Jon and {Castander}, Francisco and {Cs{\'a}bai}, Istvan and {Fukugita}, Masataka and {Ivezic}, Zeljko and {Munn}, Jeffrey A. and {Nichol}, Robert C. and {Padmanabhan}, Nikhil and {Thakar}, Aniruddha R. and {Weinberg}, David H. and {York}, Donald},
        title = "{Stellar masses and star formation histories for {}10$^{5}$ galaxies from the Sloan Digital Sky Survey}",
      journal = {\mnras},
         year = 2003,
        month = may,
       volume = {341},
       number = {1},
        pages = {33-53},
          doi = {10.1046/j.1365-8711.2003.06291.x},
archivePrefix = {arXiv},
       eprint = {astro-ph/0204055},
 primaryClass = {astro-ph},
       adsurl = {https://ui.adsabs.harvard.edu/abs/2003MNRAS.341...33K}
}

@ARTICLE{Kelly2007,
       author = {{Kelly}, Brandon C.},
        title = "{Some Aspects of Measurement Error in Linear Regression of Astronomical Data}",
      journal = {\apj},
         year = 2007,
        month = aug,
       volume = {665},
       number = {2},
        pages = {1489-1506},
          doi = {10.1086/519947},
archivePrefix = {arXiv},
       eprint = {0705.2774},
 primaryClass = {astro-ph},
       adsurl = {https://ui.adsabs.harvard.edu/abs/2007ApJ...665.1489K}
}

@ARTICLE{Kennicutt1998a,
       author = {{Kennicutt}, Robert C.},
        title = "{Star Formation in Galaxies Along the Hubble Sequence}",
      journal = {\araa},
         year = 1998,
        month = jan,
       volume = {36},
        pages = {189-232},
          doi = {10.1146/annurev.astro.36.1.189},
archivePrefix = {arXiv},
       eprint = {astro-ph/9807187},
 primaryClass = {astro-ph},
       adsurl = {https://ui.adsabs.harvard.edu/abs/1998ARA&A..36..189K}
}

@ARTICLE{Kennicutt1998b,
       author = {{Kennicutt}, Robert C.},
        title = "{The Global Schmidt Law in Star-forming Galaxies}",
      journal = {\apj},
         year = 1998,
        month = may,
       volume = {498},
       number = {2},
        pages = {541-552},
          doi = {10.1086/305588},
archivePrefix = {arXiv},
       eprint = {astro-ph/9712213},
 primaryClass = {astro-ph},
       adsurl = {https://ui.adsabs.harvard.edu/abs/1998ApJ...498..541K}
}

@ARTICLE{Kennicutt2012,
       author = {{Kennicutt}, Robert C. and {Evans}, Neal J.},
        title = "{Star Formation in the Milky Way and Nearby Galaxies}",
      journal = {\araa},
         year = "2012",
        month = "Sep",
       volume = {50},
        pages = {531-608},
          doi = {10.1146/annurev-astro-081811-125610},
archivePrefix = {arXiv},
       eprint = {1204.3552},
 primaryClass = {astro-ph.GA},
       adsurl = {https://ui.adsabs.harvard.edu/abs/2012ARA&A..50..531K}
}

@ARTICLE{Kewley2002,
       author = {{Kewley}, L.~J. and {Dopita}, M.~A.},
        title = "{Using Strong Lines to Estimate Abundances in Extragalactic H II Regions and Starburst Galaxies}",
      journal = {\apjs},
         year = 2002,
        month = sep,
       volume = {142},
       number = {1},
        pages = {35-52},
          doi = {10.1086/341326},
archivePrefix = {arXiv},
       eprint = {astro-ph/0206495},
 primaryClass = {astro-ph},
       adsurl = {https://ui.adsabs.harvard.edu/abs/2002ApJS..142...35K}
}

@ARTICLE{Kewley2004,
       author = {{Kewley}, Lisa J. and {Geller}, Margaret J. and {Jansen}, Rolf A.},
        title = "{[O II] as a Star Formation Rate Indicator}",
      journal = {\aj},
         year = 2004,
        month = apr,
       volume = {127},
       number = {4},
        pages = {2002-2030},
          doi = {10.1086/382723},
archivePrefix = {arXiv},
       eprint = {astro-ph/0401172},
 primaryClass = {astro-ph},
       adsurl = {https://ui.adsabs.harvard.edu/abs/2004AJ....127.2002K}
}

@ARTICLE{Kewley2006,
       author = {{Kewley}, Lisa J. and {Groves}, Brent and {Kauffmann}, Guinevere and {Heckman}, Tim},
        title = "{The host galaxies and classification of active galactic nuclei}",
      journal = {\mnras},
         year = 2006,
        month = nov,
       volume = {372},
       number = {3},
        pages = {961-976},
          doi = {10.1111/j.1365-2966.2006.10859.x},
archivePrefix = {arXiv},
       eprint = {astro-ph/0605681},
 primaryClass = {astro-ph},
       adsurl = {https://ui.adsabs.harvard.edu/abs/2006MNRAS.372..961K}
}

@ARTICLE{Kewley2008,
       author = {{Kewley}, Lisa J. and {Ellison}, Sara L.},
        title = "{Metallicity Calibrations and the Mass-Metallicity Relation for Star-forming Galaxies}",
      journal = {\apj},
         year = 2008,
        month = jul,
       volume = {681},
       number = {2},
        pages = {1183-1204},
          doi = {10.1086/587500},
archivePrefix = {arXiv},
       eprint = {0801.1849},
 primaryClass = {astro-ph},
       adsurl = {https://ui.adsabs.harvard.edu/abs/2008ApJ...681.1183K}
}

@ARTICLE{Kewley2019a,
       author = {{Kewley}, Lisa J. and {Nicholls}, David C. and {Sutherland}, Ralph and {Rigby}, Jane R. and {Acharya}, Ayan and {Dopita}, Michael A. and {Bayliss}, Matthew B.},
        title = "{Theoretical ISM Pressure and Electron Density Diagnostics for Local and High-redshift Galaxies}",
      journal = {\apj},
         year = 2019,
        month = jul,
       volume = {880},
       number = {1},
          eid = {16},
        pages = {16},
          doi = {10.3847/1538-4357/ab16ed},
archivePrefix = {arXiv},
       eprint = {1908.05504},
 primaryClass = {astro-ph.GA},
       adsurl = {https://ui.adsabs.harvard.edu/abs/2019ApJ...880...16K}
}

@ARTICLE{Kewley2019b,
       author = {{Kewley}, Lisa J. and {Nicholls}, David C. and {Sutherland}, Ralph S.},
        title = "{Understanding Galaxy Evolution Through Emission Lines}",
      journal = {\araa},
         year = 2019,
        month = aug,
       volume = {57},
        pages = {511-570},
          doi = {10.1146/annurev-astro-081817-051832},
archivePrefix = {arXiv},
       eprint = {1910.09730},
 primaryClass = {astro-ph.GA},
       adsurl = {https://ui.adsabs.harvard.edu/abs/2019ARA&A..57..511K}
}

@ARTICLE{Kocevski2023,
       author = {{Kocevski}, Dale D. and {Onoue}, Masafusa and {Inayoshi}, Kohei and {Trump}, Jonathan R. and {Arrabal Haro}, Pablo and {Grazian}, Andrea and {Dickinson}, Mark and {Finkelstein}, Steven L. and {Kartaltepe}, Jeyhan S. and {Hirschmann}, Michaela and {Aird}, James and {Holwerda}, Benne W. and {Fujimoto}, Seiji and {Juneau}, St{\'e}phanie and {Amor{\'\i}n}, Ricardo O. and {Backhaus}, Bren E. and {Bagley}, Micaela B. and {Barro}, Guillermo and {Bell}, Eric F. and {Bisigello}, Laura and {Calabr{\`o}}, Antonello and {Cleri}, Nikko J. and {Cooper}, M.~C. and {Ding}, Xuheng and {Grogin}, Norman A. and {Ho}, Luis C. and {Hutchison}, Taylor A. and {Inoue}, Akio K. and {Jiang}, Linhua and {Jones}, Brenda and {Koekemoer}, Anton M. and {Li}, Wenxiu and {Li}, Zhengrong and {McGrath}, Elizabeth J. and {Molina}, Juan and {Papovich}, Casey and {P{\'e}rez-Gonz{\'a}lez}, Pablo G. and {Pirzkal}, Nor and {Wilkins}, Stephen M. and {Yang}, Guang and {Yung}, L.~Y. Aaron},
        title = "{Hidden Little Monsters: Spectroscopic Identification of Low-mass, Broad-line AGNs at z > 5 with CEERS}",
      journal = {\apjl},
         year = 2023,
        month = sep,
       volume = {954},
       number = {1},
          eid = {L4},
        pages = {L4},
          doi = {10.3847/2041-8213/ace5a0},
archivePrefix = {arXiv},
       eprint = {2302.00012},
 primaryClass = {astro-ph.GA},
       adsurl = {https://ui.adsabs.harvard.edu/abs/2023ApJ...954L...4K}
}

@ARTICLE{Kriek2013,
       author = {{Kriek}, Mariska and {Conroy}, Charlie},
        title = "{The Dust Attenuation Law in Distant Galaxies: Evidence for Variation with Spectral Type}",
      journal = {\apjl},
         year = 2013,
        month = sep,
       volume = {775},
       number = {1},
          eid = {L16},
        pages = {L16},
          doi = {10.1088/2041-8205/775/1/L16},
archivePrefix = {arXiv},
       eprint = {1308.1099},
 primaryClass = {astro-ph.CO},
       adsurl = {https://ui.adsabs.harvard.edu/abs/2013ApJ...775L..16K}
}

@ARTICLE{Kroupa2001,
       author = {{Kroupa}, Pavel},
        title = "{On the variation of the initial mass function}",
      journal = {\mnras},
         year = 2001,
        month = apr,
       volume = {322},
       number = {2},
        pages = {231-246},
          doi = {10.1046/j.1365-8711.2001.04022.x},
archivePrefix = {arXiv},
       eprint = {astro-ph/0009005},
 primaryClass = {astro-ph},
       adsurl = {https://ui.adsabs.harvard.edu/abs/2001MNRAS.322..231K}
}

@ARTICLE{Laseter2024,
       author = {{Laseter}, Isaac H. and {Maseda}, Michael V. and {Curti}, Mirko and {Maiolino}, Roberto and {D'Eugenio}, Francesco and {Cameron}, Alex J. and {Looser}, Tobias J. and {Arribas}, Santiago and {Baker}, William M. and {Bhatawdekar}, Rachana and {Boyett}, Kristan and {Bunker}, Andrew J. and {Carniani}, Stefano and {Charlot}, Stephane and {Chevallard}, Jacopo and {Curtis-lake}, Emma and {Egami}, Eiichi and {Eisenstein}, Daniel J. and {Hainline}, Kevin and {Hausen}, Ryan and {Ji}, Zhiyuan and {Kumari}, Nimisha and {Perna}, Michele and {Rawle}, Tim and {Rix}, Hans-Walter and {Robertson}, Brant and {Rodr{\'\i}guez Del Pino}, Bruno and {Sandles}, Lester and {Scholtz}, Jan and {Smit}, Renske and {Tacchella}, Sandro and {{\"U}bler}, Hannah and {Williams}, Christina C. and {Willott}, Chris and {Witstok}, Joris},
        title = "{JADES: Detecting [OIII]{\ensuremath{\lambda}}4363 emitters and testing strong line calibrations in the high-z Universe with ultra-deep JWST/NIRSpec spectroscopy up to z {\ensuremath{\sim}} 9.5}",
      journal = {\aap},
         year = 2024,
        month = jan,
       volume = {681},
          eid = {A70},
        pages = {A70},
          doi = {10.1051/0004-6361/202347133},
archivePrefix = {arXiv},
       eprint = {2306.03120},
 primaryClass = {astro-ph.GA},
       adsurl = {https://ui.adsabs.harvard.edu/abs/2024A&A...681A..70L}
}

@ARTICLE{Lebouteiller2022,
       author = {{Lebouteiller}, V. and {Ramambason}, L.},
        title = "{Topological models to infer multiphase interstellar medium properties}",
      journal = {\aap},
         year = 2022,
        month = nov,
       volume = {667},
          eid = {A34},
        pages = {A34},
          doi = {10.1051/0004-6361/202243865},
archivePrefix = {arXiv},
       eprint = {2207.05657},
 primaryClass = {astro-ph.GA},
       adsurl = {https://ui.adsabs.harvard.edu/abs/2022A&A...667A..34L}
}

@INPROCEEDINGS{LeFevre2003,
       author = {{Le F{\`e}vre}, Oliver and {Saisse}, Michel and {Mancini}, Dario and {Brau-Nogue}, Sylvie and {Caputi}, Oreste and {Castinel}, Louis and {D'Odorico}, Sandro and {Garilli}, Bianca and {Kissler-Patig}, Markus and {Lucuix}, Christian and et al.},
        title = "{Commissioning and performances of the VLT-VIMOS instrument}",
    booktitle = {Instrument Design and Performance for Optical/Infrared Ground-based Telescopes},
         year = 2003,
       editor = {{Iye}, Masanori and {Moorwood}, Alan F.~M.},
       series = {Society of Photo-Optical Instrumentation Engineers (SPIE) Conference Series},
       volume = {4841},
        month = mar,
        pages = {1670-1681},
          doi = {10.1117/12.460959},
       adsurl = {https://ui.adsabs.harvard.edu/abs/2003SPIE.4841.1670L}
}

@ARTICLE{Leistedt2023,
       author = {{Leistedt}, Boris and {Alsing}, Justin and {Peiris}, Hiranya and {Mortlock}, Daniel and {Leja}, Joel},
        title = "{Hierarchical Bayesian Inference of Photometric Redshifts with Stellar Population Synthesis Models}",
      journal = {\apjs},
         year = 2023,
        month = jan,
       volume = {264},
       number = {1},
          eid = {23},
        pages = {23},
          doi = {10.3847/1538-4365/ac9d99},
archivePrefix = {arXiv},
       eprint = {2207.07673},
 primaryClass = {astro-ph.IM},
       adsurl = {https://ui.adsabs.harvard.edu/abs/2023ApJS..264...23L}
}

@ARTICLE{Leja2017,
       author = {{Leja}, Joel and {Johnson}, Benjamin D. and {Conroy}, Charlie and {van Dokkum}, Pieter G. and {Byler}, Nell},
        title = "{Deriving Physical Properties from Broadband Photometry with Prospector: Description of the Model and a Demonstration of its Accuracy Using 129 Galaxies in the Local Universe}",
      journal = {\apj},
         year = 2017,
        month = mar,
       volume = {837},
       number = {2},
          eid = {170},
        pages = {170},
          doi = {10.3847/1538-4357/aa5ffe},
archivePrefix = {arXiv},
       eprint = {1609.09073},
 primaryClass = {astro-ph.GA},
       adsurl = {https://ui.adsabs.harvard.edu/abs/2017ApJ...837..170L}
}

@ARTICLE{Leja2019,
       author = {{Leja}, Joel and {Carnall}, Adam C. and {Johnson}, Benjamin D. and
         {Conroy}, Charlie and {Speagle}, Joshua S.},
        title = "{How to Measure Galaxy Star Formation Histories. II. Nonparametric Models}",
      journal = {\apj},
         year = "2019",
        month = "May",
       volume = {876},
       number = {1},
          eid = {3},
        pages = {3},
          doi = {10.3847/1538-4357/ab133c},
archivePrefix = {arXiv},
       eprint = {1811.03637},
 primaryClass = {astro-ph.GA},
       adsurl = {https://ui.adsabs.harvard.edu/abs/2019ApJ...876....3L}
}

@ARTICLE{Lewis2026,
       author = {{Lewis}, Zach and {Maseda}, Michael V. and {de Graaff}, Anna and {Leja}, Joel and {Wang}, Bingjie and {Rix}, Hans-Walter and {McConachie}, Ian and {Cleri}, Nikko J. and {Bezanson}, Rachel and {Boogaard}, Leindert A. and {Brammer}, Gabriel and {Greene}, Jenny E. and {Hirschmann}, Michaela and {Katz}, Harley and {Labb{\'e}}, Ivo and {Matthee}, Jorryt and {Miller}, Tim B. and {Naidu}, Rohan P. and {Oesch}, Pascal A. and {Setton}, David J. and {Suess}, Katherine A. and {Weibel}, Andrea and {Whitaker}, Katherine E. and {Williams}, Christina C.},
        title = "{The Mass─Metallicity Relation and Its Observational Effects at z {\ensuremath{\sim}} 3─6}",
      journal = {\apj},
         year = 2026,
        month = jul,
       volume = {1005},
       number = {2},
          eid = {159},
        pages = {159},
          doi = {10.3847/1538-4357/ae7bfc},
archivePrefix = {arXiv},
       eprint = {2512.03134},
 primaryClass = {astro-ph.CO},
       adsurl = {https://ui.adsabs.harvard.edu/abs/2026ApJ..1005..159L}
}

@ARTICLE{Li2025,
       author = {{Li}, Yijia and {Leja}, Joel and {Johnson}, Benjamin D. and {Tacchella}, Sandro and {Davies}, Rebecca and {Belli}, Sirio and {Park}, Minjung and {Emami}, Razieh},
        title = "{Cue: A Fast and Flexible Photoionization Emulator for Modeling Nebular Emission Powered by Almost Any Ionizing Source}",
      journal = {\apj},
         year = 2025,
        month = jun,
       volume = {986},
       number = {1},
          eid = {9},
        pages = {9},
          doi = {10.3847/1538-4357/adcab4},
archivePrefix = {arXiv},
       eprint = {2405.04598},
 primaryClass = {astro-ph.GA},
       adsurl = {https://ui.adsabs.harvard.edu/abs/2025ApJ...986....9L}
}

@ARTICLE{Magg2022,
       author = {{Magg}, Ekaterina and {Bergemann}, Maria and {Serenelli}, Aldo and {Bautista}, Manuel and {Plez}, Bertrand and {Heiter}, Ulrike and {Gerber}, Jeffrey M. and {Ludwig}, Hans-G{\"u}nter and {Basu}, Sarbani and {Ferguson}, Jason W. and {Gallego}, Helena Carvajal and {Gamrath}, S{\'e}bastien and {Palmeri}, Patrick and {Quinet}, Pascal},
        title = "{Observational constraints on the origin of the elements. IV. Standard composition of the Sun}",
      journal = {\aap},
         year = 2022,
        month = may,
       volume = {661},
          eid = {A140},
        pages = {A140},
          doi = {10.1051/0004-6361/202142971},
archivePrefix = {arXiv},
       eprint = {2203.02255},
 primaryClass = {astro-ph.SR},
       adsurl = {https://ui.adsabs.harvard.edu/abs/2022A&A...661A.140M}
}

@ARTICLE{Maiolino2019,
       author = {{Maiolino}, R. and {Mannucci}, F.},
        title = "{De re metallica: the cosmic chemical evolution of galaxies}",
      journal = {\aapr},
         year = 2019,
        month = feb,
       volume = {27},
       number = {1},
          eid = {3},
        pages = {3},
          doi = {10.1007/s00159-018-0112-2},
archivePrefix = {arXiv},
       eprint = {1811.09642},
 primaryClass = {astro-ph.GA},
       adsurl = {https://ui.adsabs.harvard.edu/abs/2019A&ARv..27....3M}
}

@ARTICLE{Mannucci2010,
       author = {{Mannucci}, F. and {Cresci}, G. and {Maiolino}, R. and {Marconi}, A. and {Gnerucci}, A.},
        title = "{A fundamental relation between mass, star formation rate and metallicity in local and high-redshift galaxies}",
      journal = {\mnras},
         year = 2010,
        month = nov,
       volume = {408},
       number = {4},
        pages = {2115-2127},
          doi = {10.1111/j.1365-2966.2010.17291.x},
archivePrefix = {arXiv},
       eprint = {1005.0006},
 primaryClass = {astro-ph.CO},
       adsurl = {https://ui.adsabs.harvard.edu/abs/2010MNRAS.408.2115M}
}

@ARTICLE{Marconi2024,
       author = {{Marconi}, A. and {Amiri}, A. and {Feltre}, A. and {Belfiore}, F. and {Cresci}, G. and {Curti}, M. and {Mannucci}, F. and {Bertola}, E. and {Brazzini}, M. and {Carniani}, S. and et al.},
        title = "{HOMERUN: A new approach to photoionization modeling: I. Reproducing observed emission lines with percent accuracy and obtaining accurate physical properties of the ionized gas}",
      journal = {\aap},
         year = 2024,
        month = sep,
       volume = {689},
          eid = {A78},
        pages = {A78},
          doi = {10.1051/0004-6361/202449240},
archivePrefix = {arXiv},
       eprint = {2401.13028},
 primaryClass = {astro-ph.GA},
       adsurl = {https://ui.adsabs.harvard.edu/abs/2024A&A...689A..78M}
}

@ARTICLE{Matthee2024,
       author = {{Matthee}, Jorryt and {Naidu}, Rohan P. and {Brammer}, Gabriel and {Chisholm}, John and {Eilers}, Anna-Christina and {Goulding}, Andy and {Greene}, Jenny and {Kashino}, Daichi and {Labbe}, Ivo and {Lilly}, Simon J. and {Mackenzie}, Ruari and {Oesch}, Pascal A. and {Weibel}, Andrea and {Wuyts}, Stijn and {Xiao}, Mengyuan and {Bordoloi}, Rongmon and {Bouwens}, Rychard and {van Dokkum}, Pieter and {Illingworth}, Garth and {Kramarenko}, Ivan and {Maseda}, Michael V. and {Mason}, Charlotte and {Meyer}, Romain A. and {Nelson}, Erica J. and {Reddy}, Naveen A. and {Shivaei}, Irene and {Simcoe}, Robert A. and {Yue}, Minghao},
        title = "{Little Red Dots: An Abundant Population of Faint Active Galactic Nuclei at z {\ensuremath{\sim}} 5 Revealed by the EIGER and FRESCO JWST Surveys}",
      journal = {\apj},
         year = 2024,
        month = mar,
       volume = {963},
       number = {2},
          eid = {129},
        pages = {129},
          doi = {10.3847/1538-4357/ad2345},
archivePrefix = {arXiv},
       eprint = {2306.05448},
 primaryClass = {astro-ph.GA},
       adsurl = {https://ui.adsabs.harvard.edu/abs/2024ApJ...963..129M}
}

@INPROCEEDINGS{McLean2010,
       author = {{McLean}, Ian S. and {Steidel}, Charles C. and {Epps}, Harland and {Matthews}, Keith and {Adkins}, Sean and {Konidaris}, Nicholas and {Weber}, Bob and {Aliado}, Ted and {Brims}, George and {Canfield}, John and et al.},
        title = "{Design and development of MOSFIRE: the multi-object spectrometer for infrared exploration at the Keck Observatory}",
    booktitle = {Ground-based and Airborne Instrumentation for Astronomy III},
         year = 2010,
       editor = {{McLean}, Ian S. and {Ramsay}, Suzanne K. and {Takami}, Hideki},
       series = {Society of Photo-Optical Instrumentation Engineers (SPIE) Conference Series},
       volume = {7735},
        month = jul,
          eid = {77351E},
        pages = {77351E},
          doi = {10.1117/12.856715},
       adsurl = {https://ui.adsabs.harvard.edu/abs/2010SPIE.7735E..1EM}
}

@INPROCEEDINGS{McLean2012,
       author = {{McLean}, Ian S. and {Steidel}, Charles C. and {Epps}, Harland W. and {Konidaris}, Nicholas and {Matthews}, Keith Y. and {Adkins}, Sean and {Aliado}, Theodore and {Brims}, George and {Canfield}, John M. and {Cromer}, John L. and {Fucik}, Jason and {Kulas}, Kristin and {Mace}, Greg and {Magnone}, Ken and {Rodriguez}, Hector and {Rudie}, Gwen and {Trainor}, Ryan and {Wang}, Eric and {Weber}, Bob and {Weiss}, Jason},
        title = "{MOSFIRE, the multi-object spectrometer for infra-red exploration at the Keck Observatory}",
    booktitle = {Ground-based and Airborne Instrumentation for Astronomy IV},
         year = 2012,
       editor = {{McLean}, Ian S. and {Ramsay}, Suzanne K. and {Takami}, Hideki},
       series = {Society of Photo-Optical Instrumentation Engineers (SPIE) Conference Series},
       volume = {8446},
        month = sep,
          eid = {84460J},
        pages = {84460J},
          doi = {10.1117/12.924794},
       adsurl = {https://ui.adsabs.harvard.edu/abs/2012SPIE.8446E..0JM}
}

@ARTICLE{Miller2026,
       author = {{Miller}, Tim B. and {Zhang}, Yunchong and {Price}, Sedona H. and {Suess}, Katherine A. and {Bezanson}, Rachel and {Setton}, David J. and {Labbe}, Ivo and {Brammer}, Gabriel and {Cutler}, Sam E. and {Furtak}, Lukas J. and et al.},
        title = "{Everything Every Band All at Once II: The Relationship Between Optical Size and Stellar Mass Over Eight Billion Years of Cosmic History}",
      journal = {arXiv e-prints},
         year = 2026,
        month = mar,
          eid = {arXiv:2603.01370},
        pages = {arXiv:2603.01370},
          doi = {10.48550/arXiv.2603.01370},
archivePrefix = {arXiv},
       eprint = {2603.01370},
 primaryClass = {astro-ph.GA},
       adsurl = {https://ui.adsabs.harvard.edu/abs/2026arXiv260301370M}
}

@ARTICLE{Moreschini2026,
       author = {{Moreschini}, Bianca and {Belfiore}, Francesco and {Marconi}, Alessandro and {Cataldi}, Elisa and {Curti}, Mirko and {Amiri}, Amirnezam and {Feltre}, Anna and {Mannucci}, Filippo and {Bertola}, Elena and {Bracci}, Caterina and et al.},
        title = "{One cloud is not enough: extreme conditions bias chemical abundances in high-redshift galaxies}",
      journal = {arXiv e-prints},
         year = 2026,
        month = jan,
          eid = {arXiv:2601.08939},
        pages = {arXiv:2601.08939},
          doi = {10.48550/arXiv.2601.08939},
archivePrefix = {arXiv},
       eprint = {2601.08939},
 primaryClass = {astro-ph.GA},
       adsurl = {https://ui.adsabs.harvard.edu/abs/2026arXiv260108939M}
}

@ARTICLE{Nakajima2014,
       author = {{Nakajima}, Kimihiko and {Ouchi}, Masami},
        title = "{Ionization state of inter-stellar medium in galaxies: evolution, SFR-M$_{*}$-Z dependence, and ionizing photon escape}",
      journal = {\mnras},
         year = 2014,
        month = jul,
       volume = {442},
       number = {1},
        pages = {900-916},
          doi = {10.1093/mnras/stu902},
archivePrefix = {arXiv},
       eprint = {1309.0207},
 primaryClass = {astro-ph.CO},
       adsurl = {https://ui.adsabs.harvard.edu/abs/2014MNRAS.442..900N}
}

@ARTICLE{Nersesian2025,
       author = {{Nersesian}, Angelos and {van der Wel}, Arjen and {Gallazzi}, Anna R. and {Kaushal}, Yasha and {Bezanson}, Rachel and {Zibetti}, Stefano and {Bell}, Eric F. and {D'Eugenio}, Francesco and {Leja}, Joel and {Martorano}, Marco and {Wu}, Po-Feng},
        title = "{More is better: Strong constraints on the stellar properties of LEGA-C z {\ensuremath{\sim}} 1 galaxies with Prospector}",
      journal = {\aap},
         year = 2025,
        month = mar,
       volume = {695},
          eid = {A86},
        pages = {A86},
          doi = {10.1051/0004-6361/202452662},
archivePrefix = {arXiv},
       eprint = {2502.03021},
 primaryClass = {astro-ph.GA},
       adsurl = {https://ui.adsabs.harvard.edu/abs/2025A&A...695A..86N}
}

@software{Newville2014,
       author = {{Newville}, Matthew and {Stensitzki}, Till and {Allen}, Daniel B. and {Ingargiola}, Antonino},
        title = "{LMFIT: Non-Linear Least-Square Minimization and Curve-Fitting for Python}",
         year = 2014,
        month = sep,
          eid = {10.5281/zenodo.11813},
          doi = {10.5281/zenodo.11813},
      version = {0.8.0},
    publisher = {Zenodo},
       adsurl = {https://ui.adsabs.harvard.edu/abs/2014zndo.....11813N}
}

@ARTICLE{Ono2013,
       author = {{Ono}, Yoshiaki and {Ouchi}, Masami and {Curtis-Lake}, Emma and {Schenker}, Matthew A. and {Ellis}, Richard S. and {McLure}, Ross J. and {Dunlop}, James S. and {Robertson}, Brant E. and {Koekemoer}, Anton M. and {Bowler}, Rebecca A.~A. and et al.},
        title = "{Evolution of the Sizes of Galaxies over 7 < z < 12 Revealed by the 2012 Hubble Ultra Deep Field Campaign}",
      journal = {\apj},
         year = 2013,
        month = nov,
       volume = {777},
       number = {2},
          eid = {155},
        pages = {155},
          doi = {10.1088/0004-637X/777/2/155},
archivePrefix = {arXiv},
       eprint = {1212.3869},
 primaryClass = {astro-ph.CO},
       adsurl = {https://ui.adsabs.harvard.edu/abs/2013ApJ...777..155O}
}

@ARTICLE{Ono2023,
       author = {{Ono}, Yoshiaki and {Harikane}, Yuichi and {Ouchi}, Masami and {Yajima}, Hidenobu and {Abe}, Makito and {Isobe}, Yuki and {Shibuya}, Takatoshi and {Wise}, John H. and {Zhang}, Yechi and {Nakajima}, Kimihiko and et al.},
        title = "{Morphologies of Galaxies at z {\ensuremath{\gtrsim}} 9 Uncovered by JWST/NIRCam Imaging: Cosmic Size Evolution and an Identification of an Extremely Compact Bright Galaxy at z 12}",
      journal = {\apj},
         year = 2023,
        month = jul,
       volume = {951},
       number = {1},
          eid = {72},
        pages = {72},
          doi = {10.3847/1538-4357/acd44a},
archivePrefix = {arXiv},
       eprint = {2208.13582},
 primaryClass = {astro-ph.GA},
       adsurl = {https://ui.adsabs.harvard.edu/abs/2023ApJ...951...72O}
}

@ARTICLE{Ono2025,
       author = {{Ono}, Yoshiaki and {Ouchi}, Masami and {Harikane}, Yuichi and {Yajima}, Hidenobu and {Nakajima}, Kimihiko and {Fujimoto}, Seiji and {Nakane}, Minami and {Xu}, Yi},
        title = "{Morphological Demographics of Galaxies at z {\ensuremath{\sim}} 10{\textendash}16: Log-normal Size Distribution and Exponential Profiles Consistent with the Disk Formation Scenario}",
      journal = {\apj},
         year = 2025,
        month = oct,
       volume = {991},
       number = {2},
          eid = {222},
        pages = {222},
          doi = {10.3847/1538-4357/adfc4d},
archivePrefix = {arXiv},
       eprint = {2502.08885},
 primaryClass = {astro-ph.GA},
       adsurl = {https://ui.adsabs.harvard.edu/abs/2025ApJ...991..222O}
}

@BOOK{Osterbrock1989,
       author = {{Osterbrock}, Donald E.},
        title = "{Astrophysics of gaseous nebulae and active galactic nuclei}",
         year = 1989,
       adsurl = {https://ui.adsabs.harvard.edu/abs/1989agna.book.....O}
}

@BOOK{Osterbrock2006,
       author = {{Osterbrock}, Donald E. and {Ferland}, Gary J.},
        title = "{Astrophysics of gaseous nebulae and active galactic nuclei}",
         year = 2006,
       adsurl = {https://ui.adsabs.harvard.edu/abs/2006agna.book.....O}
}

@ARTICLE{Pagel1979,
       author = {{Pagel}, B.~E.~J. and {Edmunds}, M.~G. and {Blackwell}, D.~E. and {Chun}, M.~S. and {Smith}, G.},
        title = "{On the composition of H II regions in southern galaxies - I. NGC 300 and 1365.}",
      journal = {\mnras},
         year = 1979,
        month = oct,
       volume = {189},
        pages = {95-113},
          doi = {10.1093/mnras/189.1.95},
       adsurl = {https://ui.adsabs.harvard.edu/abs/1979MNRAS.189...95P}
}

@ARTICLE{Papovich2022,
       author = {{Papovich}, Casey and {Simons}, Raymond C. and {Estrada-Carpenter}, Vicente and {Matharu}, Jasleen and {Momcheva}, Ivelina and {Trump}, Jonathan R. and {Backhaus}, Bren E. and {Brammer}, Gabriel and {Cleri}, Nikko J. and {Finkelstein}, Steven L. and {Giavalisco}, Mauro and {Ji}, Zhiyuan and {Jung}, Intae and {Kewley}, Lisa J. and {Nicholls}, David C. and {Pirzkal}, Norbert and {Rafelski}, Marc and {Weiner}, Benjamin},
        title = "{CLEAR: The Ionization and Chemical-enrichment Properties of Galaxies at 1.1 < z < 2.3}",
      journal = {\apj},
         year = 2022,
        month = sep,
       volume = {937},
       number = {1},
          eid = {22},
        pages = {22},
          doi = {10.3847/1538-4357/ac8058},
archivePrefix = {arXiv},
       eprint = {2205.05090},
 primaryClass = {astro-ph.GA},
       adsurl = {https://ui.adsabs.harvard.edu/abs/2022ApJ...937...22P}
}

@ARTICLE{Pillepich2018a,
       author = {{Pillepich}, Annalisa and {Springel}, Volker and {Nelson}, Dylan and {Genel}, Shy and {Naiman}, Jill and {Pakmor}, R{\"u}diger and {Hernquist}, Lars and {Torrey}, Paul and {Vogelsberger}, Mark and {Weinberger}, Rainer and {Marinacci}, Federico},
        title = "{Simulating galaxy formation with the IllustrisTNG model}",
      journal = {\mnras},
         year = 2018,
        month = jan,
       volume = {473},
       number = {3},
        pages = {4077-4106},
          doi = {10.1093/mnras/stx2656},
archivePrefix = {arXiv},
       eprint = {1703.02970},
 primaryClass = {astro-ph.GA},
       adsurl = {https://ui.adsabs.harvard.edu/abs/2018MNRAS.473.4077P}
}

@ARTICLE{Pillepich2019,
       author = {{Pillepich}, Annalisa and {Nelson}, Dylan and {Springel}, Volker and {Pakmor}, R{\"u}diger and {Torrey}, Paul and {Weinberger}, Rainer and {Vogelsberger}, Mark and {Marinacci}, Federico and {Genel}, Shy and {van der Wel}, Arjen and {Hernquist}, Lars},
        title = "{First results from the TNG50 simulation: the evolution of stellar and gaseous discs across cosmic time}",
      journal = {\mnras},
         year = 2019,
        month = dec,
       volume = {490},
       number = {3},
        pages = {3196-3233},
          doi = {10.1093/mnras/stz2338},
archivePrefix = {arXiv},
       eprint = {1902.05553},
 primaryClass = {astro-ph.GA},
       adsurl = {https://ui.adsabs.harvard.edu/abs/2019MNRAS.490.3196P}
}
